\documentclass[12pt] {article}

\begin{document}

\thispagestyle{plain}
\setcounter{page}{1}

\input amssym.tex

\newcommand{\Lor}{L^{\uparrow}_{+}}

\newtheorem{lem}{Lemma}
\newtheorem{defin}{Definition}
\newtheorem{theor}{Theorem}
\newtheorem{rem}{Remark}
\newtheorem{prop}{Proposition}
\newtheorem{cor}{Corollary}
\newenvironment{demo}
{\bgroup\par\smallskip\noindent{\it Proof: }}{\rule{0.5em}{0.5em}
\egroup}

\title{Superalgebras of Dirac operators on manifolds
with special Killing-Yano tensors}

\author{Ion I. Cot\u aescu \thanks{E-mail:~~~cota@physics.uvt.ro}\\
{\small \it West University of Timi\c soara,}\\
       {\small \it V. P\^ arvan Ave. 4, RO-300223 Timi\c soara, Romania}
\and
Mihai Visinescu \thanks{E-mail:~~~mvisin@theory.nipne.ro}\\
{\small \it Department of Theoretical Physics,}\\
{\small \it National Institute for Physics and Nuclear Engineering,}\\
{\small \it P.O.Box M.G.-6, Magurele, Bucharest, Romania}}
\date{}

\maketitle

\begin{abstract}

We present the properties of new Dirac-type
operators generated by real or complex-valued special Killing-Yano tensors that
are covariantly constant and represent roots of the metric tensor. In the
real case these are just the so called complex or hyper-complex structures
of the K\" ahlerian manifolds. Such a Killing-Yano tensor produces 
simultaneously a Dirac-type operator and the generator of a one-parameter Lie 
group connecting
this operator with the standard Dirac one. In this way the Dirac operators are
related among themselves through continuous transformations associated 
with specific discrete ones. We show that the group of  these continuous 
transformations can be only $U(1)$ or $SU(2)$.  It is pointed out that the
Dirac and Dirac-type operators can form
${\cal N}=4$ superalgebras whose automorphisms combine isometries
with the $SU(2)$ transformation generated by the Killing-Yano tensors.
As an example we study the automorphisms of the superalgebras of Dirac 
operators on Minkowski spacetime.

Pacs 04.62.+v

\end{abstract}


\section{Introduction}

The quantum physics in curved backgrounds uses operators
acting on spaces of vector, tensor or spinor fields whose properties depend on
the geometry of the manifolds where these objects are defined. A crucial
problem is to find the symmetries having geometrical sources and the related
operators. The problem is not trivial since, beside the evident geometrical 
symmetry given by isometries, there are different types of hidden symmetries 
frequently associated with supersymmetries that deserve to be carefully 
studied.

The isometries are related to the existence of the Killing vectors that
give rise to the orbital operators of the scalar quantum theory commuting
with that of the free field equation. In the theories with spin these operators
get specific spin terms whose form is strongly dependent on the local
non-holonomic frames we choose by fixing the gauge \cite{CML,DGB}.
Recently the theory of isometries was extended allowing
one to pick up well-defined conserved quantities in theories with matter fields
of {\em any spin} \cite{ES,CDS}.

Another type of geometrical objects related to several specific
supersymmetries are the Killing-Yano (K-Y) tensors \cite{Y} and the
{St\" ackel-Killing} (S-K) tensors of any rank. The K-Y tensors play
an important role in theories with spin and especially in the Dirac
theory on curved spacetimes where they produce first order
differential operators, called Dirac-type operators, which
anticommute with the standard Dirac one, $D$ \cite{CML,MaCa}.
Another virtue of the K-Y tensors is that they enter as square roots
in the structure of several second rank {S-K} tensors that generate
conserved quantities in classical mechanics or conserved operators
which commute with $D$. The construction of Ref. \cite{CML} depends
upon the remarkable fact that the {S-K} tensors must have  square
root in terms of K-Y tensors in order to eliminate the quantum
anomaly and produce operators commuting with $D$ \cite{PF}. These
attributes of the K-Y tensors lead to an efficient mechanism of
supersymmetry especially when the S-K tensor is just the metric
tensor and the corresponding roots are covariantly constant K-Y
tensors.  Then each tensor of this type, $f^i$, called from now {\em
unit root}, gives rise to a Dirac-type operator, $D^i$, representing
a supercharge of a non-trivial superalgebra $\{ D^i , D^j \} \propto
D^2 \delta_{ij}$ \cite{CV2}. The real-valued unit roots are nothing
other than the complex or the hyper-complex structures of the K\"
ahlerian manifolds. However, it was shown that operators $D^i$  can
be produced by unit roots with complex-valued components
\cite{K1,K2}. This represents an extension of the K\" ahlerian
geometries that seems to be productive for the Dirac theory since it
permits to construct superalgebras of Dirac-type operators even on
the Minkowski spacetime which is not K\" ahlerian, having only
complex-valued unit roots \cite{K2,CV7}.

It is known that in four-dimensional manifolds the standard Dirac
operator and the Dirac-type ones can be related among themselves
through continuous or discrete transformations \cite{CV6,K2}. It is
interesting that there are only two possibilities, namely either
transformations of the $U(1)$ group or $SU(2)$ transformations
\cite{CV6,K2,CV7}. In the case of the real-valued unit roots, the
first type of symmetry is proper to  K\" ahler manifolds while the
second largest one is characteristic for hyper-K\" ahler geometries
\cite{CV6}. However,  similar results can be obtained also in the
non-K\" ahlerian case of  complex-valued unit roots. Recently we
have shown that the symmetries of this type can not be larger than
$SU(2)$ \cite{CV7}. Moreover, we pointed out that their
transformations can be mixed with the isometries in order to obtain
the automorphisms of the superalgebras of Dirac and Dirac-type
operators of the hyper-K\" ahler manifolds \cite{NOVA}. In the
present paper we continue this study focusing on the automorphisms
of the superalgebras of Dirac operators on non-K\" ahlerian
manifolds, including the Minkowski spacetime.

The paper is organized as a report including the previous results
one needs for presenting the new topic. We start in the second
section with the construction of a simple version of the Dirac
theory in manifolds of any dimensions. In the next section we
present the theory of external symmetry of the Dirac field
\cite{ES}. The unit roots are defined in the fourth section giving
their main properties in the K\" alerian case as well as in the
non-K\" ahlerian one. The main conclusion is that there are either
single unit roots or triplets of unit roots which have special
algebraic properties similar to those of the quaternion algebra. The
Dirac-type operators produced by unit roots  are introduced in the
next section showing that these and the standard one, $D$, can be
organized as superalgebras, with ${\cal N}=2$ in the case of single
unit roots and ${\cal N}=4$ when one has triplets. The sixth section
is devoted to the continuous symmetries generated by unit roots that
relate the Dirac operators among themselves. We show that these are
given by the group $U(1)$ when ${\cal N}=2$ or by the group $SU(2)$
in the case of ${\cal N}=4$. In the next one we study the effects of
isometries  pointing out that the isolated unit roots are invariant
while the triplets transform according to a specific induced
representation of the isometry group. Using these elements we
discuss the automorphisms of our superalgebras in the section eight.
Finally we study the superalgebras of the Dirac and Dirac-type
operators of the Minkowski spacetime.

\section{The Dirac theory in any dimensions}

The gauge-covariant theory of the Dirac field can be formulated in
any (non-holonomic) orthogonal local frame of a (pseudo-)Riemannian manifold
$M_n$. Thus we must consider simultaneously a local frame and an usual
natural frame represented by the local chart where we introduce
the coordinates, $x^{\mu}$, $\mu,\nu,...=1,2,...n$. The local frames
are defined by the gauge fields (or "vilbeins") $e(x)$ and $\hat
e(x)$, whose components are labeled by the local indices
$\hat\alpha,...,\hat\mu,...=1,2,...n$. We remind the reader that the local 
indices
are raised or lowered by the (pseudo-)Euclidean metric $\eta$ of the
flat model of $M_n$ while for the natural indices we have to use the
metric tensor $g(x)$. The fields $e$ and $\hat e$ accomplish the conditions
$e_{\hat\alpha}^{\mu}\hat e_{\nu}^{\hat\alpha}=\delta_{\mu}^{\nu}$,
$e_{\hat\alpha}^{\mu}\hat e_{\mu}^{\hat\beta}=\delta_{\hat\alpha}^{\hat\beta}$,
orthogonality relations,
$g_{\mu\nu} e_{\hat\alpha}^{\mu} e^{\nu}_{\hat\beta}=
\eta_{\hat\alpha \hat\beta}$, and give the metric tensor of $M_n$ as
$g_{\mu\nu}=\eta_{\hat\alpha \hat\beta}
\hat e ^{\hat\alpha}_{\mu} \hat e^{\hat\beta}_{\nu}$.

In what follows we shall focus only on the manifolds with an even
number of dimensions. In addition, we assume that the metric $\eta$ has
an arbitrary signature, $(n_+,n_-)$ with $n_+ +n_- =n=2k$.

\subsection{Clifford algebra and the gauge group}

For the Dirac theory in manifolds $M_n$ with $n=2k$ a
$2k+1$-dimensional Clifford algebra \cite{Clif} is enough. This is
defined on  the $2^k$-dimensional space $\Psi$ of the complex
spinors $\psi=\tilde\varphi_1 \otimes \tilde\varphi_2 ...
\otimes\tilde\varphi_k$ built using  complex two-dimensional Pauli
spinors $\tilde\varphi$. According to our previous general results
\cite{CV7}, we know that one can define a set of point-independent
gamma matrices $\gamma^{\hat\mu}$, labeled by local indices, such
that
\begin{equation}\label{ACOM}
\{ \gamma^{\hat\alpha},\, \gamma^{\hat\beta} \}
=2\tilde\eta^{\hat\alpha\hat\beta}{\bf 1} \,.
\end{equation}
where ${\bf 1}$ is the identity matrix.
In this form, the first $ n_{+}$ matrices $\gamma^{\hat\mu}$ remain
hermitian while the
$ n_{-}$ last ones become anti-hermitian. The unitaryness can be restored
replacing the usual Hermitian adjoint with the generalized Dirac adjoint
\cite{CV7}.
\begin{defin}
We say that $\overline{\psi}=\psi^{+}\gamma$ is the generalized Dirac adjoint
of the field $\psi$ if the hermitian matrix  $\gamma=\gamma^{+}$  satisfies
the condition $(\gamma)^2={\bf 1}$ and all the matrices $\gamma^{\hat\mu}$ are either
self-adjoint or anti self-adjoint with respect to this operation,
i.e. $\overline{\gamma}^{\hat\mu}=\gamma (\gamma^{\hat\mu})^{+}\gamma=\pm \gamma^{\hat\mu}$.
\end{defin}
It is clear that the matrix $\gamma$ play here the role of {\em metric
  operator} giving the generalized Dirac adjoint of any square matrix $X$ as
$\overline{X}=\gamma X^{+} \gamma$.

The gauge group $G(\eta)=O(n_{+}, n_{-})$ of $M_n$ defining the principal
fiber bundle is a pseudo-orthogonal group that
admits an universal covering group ${\bf G}(\eta)$ which is simply
connected and has the same Lie algebra we denote by ${\bf g}(\eta)$.
The group ${\bf G}(\eta)$ is the model of the spinor fiber
bundle that completes the spin structure we need. In order to avoid
complications due to the presence of these two groups we consider here that
the basic piece is the group ${\bf G}(\eta)$, denoting by $[\omega]$
their elements in the standard {\em covariant} parametrization given by the
skew-symmetric real parameters $\omega_{\hat\mu\hat\nu}=
-\omega_{\hat\nu\hat\mu}$. Then the
identity element of ${\bf G}(\eta)$ is $1=[0]$ and the inverse of
$[\omega]$ with respect to the group multiplication reads
$[\omega]^{-1}=[-\omega]$.
\begin{defin}
We say that the gauge group is the {\em vector} representation of
${\bf G}(\eta)$ and denote $G(\eta)=
vect[{\bf G}(\eta)]$.
The representation $spin[{\bf G}(\eta)]$ carried by the space
$\Psi$ and generated by the spin operators
\begin{equation}\label{SAB}
S^{\hat\alpha\hat\beta}=\frac{i}{4}\left[\gamma^{\hat\alpha},\,
\gamma^{\hat\beta}\right]
\end{equation}
is called the {\em spinor} representation of ${\bf G}(\tilde\eta)$. The
spin operators are the basis generators of the spinor representation
$spin[{\bf g}(\eta)]$ of the Lie algebra ${\bf g}(\eta)$.
\end{defin}
In general, the spinor representation is reducible. Its generators are
self-adjoint, $\overline{S}^{\hat\alpha\hat\beta} =S^{\hat\alpha \hat\beta}$,  
and satisfy
\begin{eqnarray}
[S^{\hat\alpha\hat\beta},\,\gamma^{\hat\sigma}]&=&
i(\eta^{\hat\beta\hat\sigma}\gamma^{\hat\alpha}-
\eta^{\hat\alpha\hat\sigma}\gamma^{\hat\beta})\,,\label{Sgg}\\
{[} S_{\hat\alpha\hat\beta},\,S_{\hat\sigma\hat\tau} {]}&=&i(
\eta_{\hat\alpha\hat\tau}\,S_{\hat\beta\hat\sigma}-
\eta_{\hat\alpha\hat\sigma}\,S_{\hat\beta\hat\tau}+
\eta_{\hat\beta\hat\sigma}\,S_{\hat\alpha\hat\tau}-
\eta_{\hat\beta\hat\tau}\,S_{\hat\alpha\hat\sigma})\,,\label{SSS}
\end{eqnarray}
as it results from Eqs. (\ref{ACOM}) and (\ref{SAB}). It is obvious that
Eq. (\ref{SSS}) gives just the canonical commutation rules of a Lie algebra
isomorphic with that of the groups $G(\eta)$ or ${\bf G}(\eta)$ \cite{CV7,NOVA}.
The spinor and vector representations are related between themselves through
the following
\begin{theor}
For any real or complex valued skew-symmetric tensor
$\omega_{\hat\alpha\hat\beta}=-\omega_{\hat\beta\hat\alpha}$ the matrix
\begin{equation}\label{TeS}
T(\omega)=e^{-iS(\omega)}\,,\quad S(\omega)=\frac{1}{2}
\omega_{\hat\alpha\hat\beta} S^{\hat\alpha\hat\beta}\,,
\end{equation}
transforms the gamma-matrices according to the rule
\begin{equation}\label{TgT}
[T(\omega)]^{-1}\gamma^{\hat\alpha}T(\omega)=\Lambda^{\hat\alpha\,\cdot}
_{\cdot\,\hat\beta}(\omega)\gamma^{\hat\beta}\,,
\end{equation}
where
\begin{equation}\label{Lam}
\Lambda^{\hat\alpha\,\cdot}_{\cdot\,\hat\beta}(\omega)=
\delta^{\hat\alpha}_{\hat\beta}
+\omega^{\hat\alpha\,\cdot}_{\cdot\,\hat\beta}
+\frac{1}{2}\,\omega^{\alpha\,\cdot}_{\cdot\,\hat\mu}
\omega^{\hat\mu\,\cdot}_{\cdot\,\hat\beta}+...
+\frac{1}{n!}\,\underbrace{\omega^{\hat\alpha\,\cdot}_{\cdot\,\hat\mu}
\omega^{\hat\mu\,\cdot}_{\cdot\, \hat\nu}
...\omega^{\hat\sigma\,\cdot}_{\cdot\,\hat\beta}}_{n}+...\,.
\end{equation}
\end{theor}
\begin{demo}
All these results can be obtained using Eqs. (\ref{ACOM}) and (\ref{SAB}).
\end{demo}\\
The real components $\omega_{\hat\alpha\hat\beta}$ are the parameters of the
covariant basis of
the Lie algebra ${\bf g}(\eta)$ giving all the  transformation
matrices $T(\omega) \in spin[{\bf G}(\eta)]$ and
$\Lambda(\omega)\in vect[{\bf G}(\eta)]$. Hereby we see that the
spinor representation $spin[{\bf G}(\eta)]$ is unitary since for
$\omega \in {\Bbb R}$ the generators $S(\omega)\in spin[{\bf g}(\eta)]$
are self-adjoint, $\overline{S}(\omega)=S(\omega)$, and the matrices
$T(\omega)$ are unitary with respect to the Dirac adjoint satisfying
$\overline{T}(\omega)=[T(\omega)]^{-1}$.

The covariant parameters $\omega$ can also take complex values.  Then this
parametrization spans the {\em complexified} group of ${\bf G}( \eta)$,
denoted by ${\bf G}_c(\eta)$, and the corresponding vector and
(non-unitary) spinor representations. Obviously, in this case the Lie
algebra is the complexified algebra ${\bf g}_c(\eta)$.
We note that from the mathematical point of view
$G(\eta)=vect[{\bf G}( \eta)]$ is the group of automorphisms of
the tangent
fiber bundle ${\cal T}(M_n)$ of $M_n$ while the transformations of
$G_c(\eta)=vect[{\bf G}_c( \eta)]$ are automorphisms of the
complexified tangent fiber bundle ${\cal T}(M_n)\otimes {\Bbb C}$.

\subsection{The Dirac field}

With these preparations, the gauge-covariant theory of the Dirac field $\psi$
can be formulated starting with the standard Dirac gauge invariant
action \cite{CV7}.  This gives the
Dirac equation $D\psi=m_0\psi$ where  $m_0$ is the fermion mass and
the Dirac operator has the form
\begin{equation}\label{Dirac}
D=i\gamma^{\mu}\nabla_{\mu}\,,\quad
\gamma^{\mu}(x)=e^{\mu}_{\hat\alpha}(x)\gamma^{\hat\alpha}\,.
\end{equation}
The covariant derivatives  
$\nabla_{\mu}=\hat e_{\mu}^{\hat\alpha}\nabla_{\hat\alpha}=
\tilde\nabla_{\mu}+\Gamma_{\mu}^{spin}$
are formed by the usual ones, $\tilde\nabla_{\mu}$
(acting in natural indices), and the spin connection
\[
\Gamma_{\mu}^{spin}=\frac{i}{2}
e^{\beta}_{\hat\nu}
(\hat e^{\hat\sigma}_{\alpha}\Gamma^{\alpha}_{\beta\mu}-
\hat e^{\hat\sigma}_{\beta,\mu} )
S^{\hat\nu\,\cdot}_{\cdot\,\hat\sigma}\,,
\]
giving $\nabla_{\mu}\psi=(\partial_{\mu}+\Gamma_{\mu}^{spin})\psi$.

Now it is obvious that our definition of the generalized Dirac
adjoint is correct since $\overline{\gamma^{\mu}}=\gamma^{\mu}$ and
$\overline{\Gamma}_{\mu}^{spin}=-\Gamma_{\mu}^{spin}$
such that the Dirac operator results to be self-adjoint, $\overline{D}=D$.
Moreover,  the quantity $\overline{\psi}\psi$ has to be derived as a
scalar, i.e. $\nabla_{\mu}(\overline{\psi}\psi)=
\overline{\nabla_{\mu}\psi}\,\psi+\overline{\psi}\,\nabla_{\mu}\psi=
\partial_{\mu}(\overline{\psi}\psi)$,
while the quantities $\overline{\psi}\gamma^{\alpha}\gamma^{\beta}...\psi$
behave as tensors of different ranks.

Thus we reproduce the main features of the familiar tetrad gauge covariant
theories with spin in (1+3)-dimensions from which we can take over all the
results arising from similar formulas. In this way we find that
the point-dependent matrices $\gamma^{\mu}(x)$ and
$S^{\mu\nu}(x)=e^{\mu}_{\hat\alpha}(x)e^{\nu}_{\hat\beta}(x)
S^{\hat\alpha\hat\beta}$ have similar properties as (\ref{ACOM}), (\ref{SAB}),
(\ref{Sgg}) and (\ref{SSS}), but written in natural indices and with $g(x)$
instead of the flat metric $\eta$. Using this algebra and the standard 
notations for
the Riemann-Christoffel curvature tensor,  $R_{\alpha\beta \mu\nu}$,
Ricci tensor, $R_{\alpha\beta}=R_{\alpha \mu \beta\nu}g^{\mu\nu}$, and scalar
curvature, $R=R_{\mu\nu}g^{\mu\nu}$, we  recover the useful formulas
\begin{eqnarray}
&&\nabla_{\mu}(\gamma^{\nu}\psi)=\gamma^{\nu}\nabla_{\mu}\psi\,,
\label{Nabg}\\
&&[\nabla_{\mu},\,\nabla_{\nu}]\psi=
\textstyle\frac{1}{4}
R_{\alpha\beta\mu\nu}\gamma^\alpha\gamma^{\beta}\psi \,,
\end{eqnarray}
and the identity $R_{\alpha\beta\mu\nu}\gamma^{\beta}
\gamma^{\mu}\gamma^{\nu}=-2R_{\alpha\nu}\gamma^{\nu}$
that allow one to calculate
\[
D^2=-\nabla^2+\textstyle{\frac{1}{4}}R\,{\bf 1}\,, \quad
\nabla^2=g^{\mu\nu}\nabla_{\mu}\nabla_{\nu}\,.
\]
It remains to complete the operator algebra with new observables from which
we have to select complete sets of commuting observables for defining quantum
modes. These can be obtained as conserved operators associated to
the isometries of $M_n$.

\section{The external symmetry of the Dirac theory}

The relativistic covariance  in the sense of general relativity is
too general to play the same role as the Lorentz or Poincar\' e covariance in
special relativity  \cite{W}. In other respects, the gauge covariance of
the theories with spin represents another kind of general symmetry that is not
able to produce itself conserved observable  \cite{SW}. Therefore, if
we look for sources of symmetries able to generate conserved quantities, we
have to concentrate mainly on isometries that point out the spacetime
symmetry  giving us the specific Killing vectors  \cite{SW,WALD,ON}. The
physical fields minimally coupled with the gravitational one take over
this symmetry, transforming according to different representations of the
isometry group. In the case of the scalar vector or tensor fields these
representations are completely defined by the well-known  rules of the
general coordinate transformations  since the isometries are in fact
particular coordinate transformations. However, the behavior under isometries
of the fields with half integer spin is more complicated since their
transformation rules explicitly depend on the gauge fixing. The specific
theory of this type of transformations is the recent theory of external
symmetry we present in this section  \cite{ES}.

\subsection{The gauge and relativistic covariance}

The use of the covariant derivatives assures the covariance of the whole theory
under the  gauge transformations,
\begin{eqnarray}
\hat e^{\hat\alpha}_{\mu}(x)&\to& \hat e'^{\hat\alpha}_{\mu}(x)=
\Lambda^{\hat\alpha\,\cdot}_{\cdot\,\hat\beta}[A(x)]
\,\hat e^{\hat\beta}_{\mu}(x)\,,\nonumber\\
e_{\hat\alpha}^{\mu}(x)&\to&  {e'}_{\hat\alpha}^{\mu}(x)=
\Lambda_{\hat\alpha\,\cdot}^{\cdot\,\hat\beta}[A(x)]
\,e_{\hat\beta}^{\mu}(x)\,,\label{gauge}\\
\psi(x)&\to&\psi'(x)=T [A(x)]\,\psi(x)\,,\nonumber
\end{eqnarray}
produced by the mappings $A: M_n\to {\bf G}(\eta)$ the values of which
are {\em local} transformations $A(x)=[\omega(x)]\in {\bf G}(\eta)$
determined by the set of {\em real} functions $\omega_{\hat\mu\hat\nu}=
-\omega_{\hat\nu\hat\mu}$ defined on $M_n$. In other words $A$ denotes
sections of the spinor fiber bundle that can be organized as
a group, ${\rm Sec}(M_n)$, with respect to the multiplication $\times$ defined
as $(A'\times A)(x)=A'(x)A(x)$. We use the notations $Id$ for the mapping
identity, $Id(x)=1\in {\bf G}(\eta)$, and $A^{-1}$ for the inverse of
$A$ which satisfies $(A^{-1})(x)=[A(x)]^{-1}$.

The general gauge-covariant theory of the Dirac spinors  must
be also covariant under  general coordinate transformation of $M_n$ which, in
the {\em passive} mode,
\footnote{We prefer the term of coordinate transformation instead of
diffeomorphism since we adopt this viewpoint.}
can be seen as changes of the local charts corresponding to the same domain of
$M_n$  \cite{WALD,ON}. If $x$ and $x'$ are the coordinates of a given point in
two different charts then there is a mapping $\phi$ between these charts giving
the coordinate transformation $x\to x'=\phi(x)$. These transformations form
the group ${\rm Sec}(M_n)$ with respect to the composition of mappings,
$\,\circ\,$, defined as usual, i.e. $(\phi'\circ\phi)(x)=\phi'[\phi(x)]$. We
denote the identity map of this group by $id$ and the inverse mapping of
$\phi$ by $\phi^{-1}$.

The coordinate transformations change all the components carrying  natural
indices including those of the gauge fields \cite{SW} changing thus the
positions of the local frames with respect to the natural ones. If we assume
that the physical experiment makes reference to the axes of the local frame
then it could appear situations when several correction of the positions of
these frames should be needed before (or after) a general coordinate
transformation. Obviously, these have to be made with the help of suitable
gauge transformations associated to the coordinate ones.
\begin{defin}
The {\em combined} transformation $(A,\phi)$ is the gauge transformation given
by the section $A\in {\rm Sec}(M_n)$ followed by the coordinate transformation
$\phi\in {\rm Aut}(M_n)$.
\end{defin}
In this new notation the pure gauge transformations appear as $(A,id)\in
{\rm Sec}(M_n)$ while the coordinate transformations will be denoted from now
by $(Id,\phi)\in {\rm Aut}(M_n)$.
The  effect of a combined transformation $(A,\phi)$ upon our basic elements
is, $x\to x'=\phi(x)$,
\begin{equation}\label{genTP}
\psi(x)\to \psi'(x')=T[A(x)]\psi(x)\,,
\end{equation}
and $e(x)\to \hat e'(x')$ where $e'$ are the transformed gauge fields  of the
components
\begin{equation}\label{eeee}
e'^{\mu}_{\hat\alpha}[\phi(x)]=\Lambda^{\cdot\,\hat\beta}
_{\hat\alpha\,\cdot}[A(x)]\,e^{\nu}_{\hat\beta}(x)
\frac{\partial\phi^{\mu}(x)}{\partial x^{\nu}}\,,
\end{equation}
while the components of $\hat e'$ have to be calculated according to Eqs.
(\ref{eeee}). Thus we have written down the most general transformation laws
that leave the theory invariant.

The association among the transformations of the gauge group  and coordinate
transformation  leads to a new group with a specific multiplication. In order
to find how looks this new operation  it is convenient to use the composition
among the mappings $A$ and $\phi$ (taken only in this order) giving the new
mappings $A\circ\phi$ defined as $(A\circ \phi)(x)=A[\phi(x)]$. The calculation
rules $Id\circ \phi=Id$, $A\circ id=A$ and
$(A'\times A)\circ \phi=(A'\circ \phi)\times (A\circ \phi)$ are obvious.
In this context one can demonstrate
\begin{theor}
The set of combined transformations of $M_n$, ${\cal G}(M_n)$,
form a group with respect to the multiplication $*$ defined as
\begin{equation}\label{comp}
(A',\phi')*(A,\phi)=
\left((A'\circ\phi)\times A,\phi'\circ\phi\right)\,.
\end{equation}
\end{theor}
\begin{demo}
First of all we observe that  $(A,\phi)=(Id,\phi)*(A,id)$.
Furthermore,  one can verify the result calculating the effect of
this product upon the field $\psi$.
\end{demo}\\
Now the identity is $(Id,id)$  while the inverse of a pair $(A,\phi)$ reads
\begin{equation}\label{compAphi}
(A,\phi)^{-1}=(A^{-1}\circ\phi^{-1},\phi^{-1})\,.
\end{equation}
In addition, one can demonstrate that the group of combined
transformations is the semi-direct product ${\cal G}(M_n)={\rm
Sec}(M_n)  \,\circledS\, {\rm Aut}(M_n)$ between the group of
sections which plays the role of invariant subgroup and that of
coordinate transformations \cite{ES}. The same construction but
starting with the group ${\bf G}_c(\eta)$ instead of ${\bf G}(\eta)$
yields the complexified group  of combined transformations, ${\cal
G}_c(M_n)$.

The use of combined transformations is justified only in theories
where there are physical reasons requiring to use local frames since
in natural frames the effect of the combined transformations on the
vector and tensor fields reduces to that of coordinate
transformations. However, the physical systems involving spinors can
be described exclusively in local frames where our theory is
essential.
\begin{defin}
The spinor representation of ${\cal G}(M_n)$ has values in the space of
the linear operators ${U}: \Psi\to \Psi$ such that for each $(A,\phi)$ there
exists an operator ${U}(A,\phi)\in spin[{\cal G}(M_n)]$ having the action
\[
{U}(A,\phi)\psi=[T(A)\psi]\circ \phi^{-1}=[T(A\circ \phi^{-1})]
(\psi\circ \phi^{-1})\,.
\]
\end{defin}
This rule gives the transformations (\ref{genTP}) in each point of $M_n$ if we
put $\psi'={U}(A,\phi)\psi$ and then calculate the value of $\psi'$ in the
point $x'=\phi(x)$.  The Dirac operator $D$ {\em covariantly} transforms as
\[
(A,\phi)~ :~~ D(x) \to D'(x')=
T[A(x)]D(x)\overline{T}[A(x)] \,,
\]
where $D'={U}(A,\phi)D[{U}(A,\phi)]^{-1}$. In general, the combined
transformations change the form of the Dirac operator which depends on the
gauge one uses ($D'\not=D$). We note that for the gauge transformations with
$\phi=id$ (when $x'=x$) the action of $U(A,id)$ reduces to the linear
transformation given by the matrix $T(A)\in spin[{\bf G}(\eta)]$.

\subsection{Isometries and the external symmetry}

In general, the symmetry of the manifold $M_n$ is given by its isometry group,
$I(M_n)\subset {\rm Aut}(M_n)$, whose transformations, $x\to x'(x)$, are
coordinate transformation which leave the metric tensor invariant in any chart
 \cite{SW,WALD,ON},
\begin{equation}\label{giso}
g_{\alpha\beta}(x')\frac{\partial x'^\alpha}{\partial x^\mu}
\frac{\partial x'^\beta}{\partial x^\nu}=g_{\mu\nu}(x)\,.
\end{equation}
The isometry group is formed by sets of coordinate transformations,
$x\to x'=\phi_{\xi}(x)$, depending on $N$ independent real parameters, $\xi^a$
($a,b,c...=1,2,...,N$), such that $\xi=0$ corresponds to the identity map,
$\phi_{\xi=0}=id$. These transformations form a Lie group equipped with  the
composition rule
\begin{equation}\label{compphi}
\phi_{\xi'}\circ \phi_{\xi}=\phi_{p(\xi',\xi)}\,,
\end{equation}
where the functions $p$  define the group multiplication. These  satisfy
$p^{a}(0,\xi)=p^{a}(\xi,0)=\xi^{a}$ and
$p^{a}(\xi^{-1},\xi)=p^{a}(\xi,\xi^{-1})=0$ where $\xi^{-1}$ are the
parameters of the inverse mapping of $\phi_{\xi}$,
$\phi_{\xi^{-1}}=\phi^{-1}_{\xi}$. Moreover, the functions $p$ give the
structure constants $c_{abc}$ of this group \cite{HAM}
which define the commutation relations of the basis generators of the Lie
algebra $i(M_n)$ of $I(M_n)$. For small values of the
group parameters the infinitesimal transformations,
$x^{\mu}\to x'^{\mu}=x^{\mu}+\xi^{a}k_{a}^{\mu}(x)+\cdots$,
are given by the Killing vectors $k_{a}$ whose components,
\begin{equation}\label{ka}
k_{a}^{\mu}=\left.{\frac{\partial \phi^{\mu}_{\xi}}
{\partial\xi^{a}}}\right|_{\xi=0}\,,
\end{equation}
satisfy the Killing equations  $k_{a\, (\mu;\nu)}\equiv k_{a\, \mu;\nu}+
k_{a\, \nu;\mu}=0$ and the identities
\begin{equation}\label{kkc}
k^{\mu}_{a}k^{\nu}_{b,\mu}
-k^{\mu}_{b}k^{\nu}_{a,\mu}+c_{abc}k^{\nu}_{c}=0\,.
\end{equation}

The simplest representation of $I(M_n)$ is the {\em natural} one carried by the
space of the {\em scalar} fields $\vartheta$ which transform as
$\vartheta \to \vartheta' =\vartheta\circ\phi_{\xi}^{-1}$. This rule defines
the operator-valued representation of the group $I(M_n)$ generated by the
operators,
\[
L_{a}=-ik_{a}^{\mu}\partial_{\mu}\,, \quad a=1,2,...,N\,,
\]
which are completely determined by the Killing vectors. From Eq. (\ref{kkc})
we see that they obey the commutation rules
\[
[L_{a}, L_{b}]=ic_{abc}L_{c}\,,
\]
given by the structure constants of the Lie algebra $i(M_n)$.

In the theories involving fields with spin, an isometry can change the
relative positions of the local frames with respect to the natural ones.
This fact may be an impediment when one intends to study the symmetries of
these theories in local frames. For this reason it is natural to
suppose that the good symmetry transformations we need are  isometries
preceded by appropriate gauge transformations which should assure that not
only the form of the metric tensor would be conserved but the form of the
gauge field components too. However, these transformations are nothing other
than {\em particular} combined transformations whose coordinate
transformations are isometries.
\begin{defin}
The {\em external symmetry} transformations, $(A_{\xi},\phi_{\xi})$,
are particular combined transformations involving isometries,
$(Id,\phi_{\xi})\in I(M_n)$, and corresponding gauge transformations,
$(A_{\xi}, id)\in {\rm Sec}(M_n)$, necessary to {\em preserve the gauge}.
\end{defin}
This requirement is accomplished only if we assume that, for  given
gauge fields $e$ and $\hat e$, $A_{\xi}$  is defined by
\begin{equation}\label{Axx}
\Lambda^{\hat\alpha\,\cdot}_{\cdot\,\hat\beta}[A_{\xi}(x)]=
\hat e_{\mu}^{\hat\alpha}[\phi_{\xi}(x)]\frac{\partial \phi^{\mu}_{\xi}(x)}
{\partial x^{\nu}}\,e^{\nu}_{\hat\beta}(x)\,,
\end{equation}
with the supplementary condition $A_{\xi=0}(x)= 1\in {\bf G}(\eta)$.
Since $\phi_{\xi}$ is an isometry Eq. (\ref{giso}) guarantees that
$\Lambda[A_{\xi}(x)]\in vect[{\bf G}(\eta)]$ and, implicitly,
$A_{\xi}(x)\in {\bf G}(\eta)$. Then the transformation laws of our fields are
\begin{equation}\label{es}
(A_{\xi},\phi_{\xi}):\qquad
\begin{array}{rlrcl}
x&\to&x'&=&\phi_{\xi}(x)\\
e(x)&\to&e'(x')&=&e[\phi_{\xi}(x)]\\
\hat e(x)&\to&\hat e'(x')&=&\hat e[\phi_{\xi}(x)]\\
\psi(x)&\to&\psi'(x')&=&T[A_{\xi}(x)]\psi(x)\,.
\end{array}
\qquad
\end{equation}
The mean virtue of these transformations is that they leave {\em invariant}
the form of the Dirac operator, $D'=D$.
\begin{theor}
The set of the external symmetry transformations $(A_{\xi},\phi_{\xi})$ form
the Lie group $S(M_n)\subset {\cal G}(M_n)$ with respect to the
operation $*$. This group, will be called the group of external symmetry
of $M_n$.
\end{theor}
\begin{demo}
Starting with Eq. (\ref{Axx})  we find that
$(A_{\xi'}\circ\phi_{\xi})\times A_{\xi}=A_{p(\xi',\xi)}$,
and, according to Eqs. (\ref{comp}) and (\ref{compphi}), we obtain
\begin{equation}\label{mult}
(A_{\xi'},\phi_{\xi'})*(A_{\xi},\phi_{\xi})=
(A_{p(\xi',\xi)},\phi_{p(\xi',\xi)})\,,
\end{equation}
and $(A_{\xi=0},\phi_{\xi=0})=(Id,id)$.
\end{demo}\\
From Eq. (\ref{mult}) we understand that $S(M_n)$ is {\em locally
isomorphic} with $I(M_n)$ and, therefore, the Lie algebra $s(M_n)$
of the group $S(M_n)$ is isomorphic with  $i(M_n)$ having the same
structure constants.  There are arguments that the group $S(M_n)$ is
isomorphic with the universal covering group of $I(M_n)$ since it
has  anyway the topology induced by ${\bf G}(\eta)$ which is simply
connected. In general, the number of group parameters of $I(M_n)$ or
$S(M_n)$ (which is equal to the number of the independent Killing
vectors of $M_n$) can be $0\le N\le \frac{1}{2}n(n+1)$ \cite{SW}.

The last of Eqs. (\ref{es}) giving the transformation law of the field
$\psi$ defines the operator-valued representation
$(A_{\xi},\phi_{\xi})\to {U}_{\xi}\equiv U(A_{\xi},\phi_{\xi})$ of the group
$S(M_n)$,
\begin{equation}\label{rep}
({U}_{\xi}\psi)[\phi_{\xi}(x)]=T[A_{\xi}(x)]\psi(x)\,,
\end{equation}
which is the spinor representation
$spin[S(M_n)]\subset spin[{\cal G}(M_n)]$ of the group $S(M_n)$.
This representation has unitary transformation matrices in the sense of the
Dirac adjoint ($\overline{T}=T^{-1}$) and its transformations leaves the
operator $D$ invariant,
\[
{U}_{\xi}D{U}_{\xi}^{-1}=D\,.
\]
Since $A_{\xi}(x)\in {\bf G}(\eta)$ we say that  $spin[S(M_n)]$ is
{\em induced} by the representation $spin[{\bf G}(\eta)]$  \cite{BR,MAK}.
\begin{theor}
The basis generators of the spinor representation $spin[s(M_n)]$ of the Lie
algebra $s(M_n)$ are \cite{CML,ES}
\begin{equation}\label{X}
X_{a}=-ik^{\mu}_{a}\nabla_{\mu}+\frac{1}{2}\,
k_{a\, \mu;\nu}\,e^{\mu}_{\hat\alpha}\,e^{\nu}_{\hat\beta}\,
S^{\hat\alpha\hat\beta} \,.
\end{equation}
\end{theor}
\begin{demo}
The proof is given in \cite{ES} where we recover
the important result of \cite{CML} derived for the Dirac field in $M_4$.
\end{demo}

\begin{theor}
The operators (\ref{X}) are self-adjoint with respect to the Dirac adjoint and
satisfy the commutation rules
\begin{equation}\label{comX}
[X_{a}, X_{b}]=ic_{abc}X_{c}\,, \quad
[D,X_{a}]=0\,, \quad a,b...=1,2,...,N\,,
\end{equation}
where $c_{abc}$ are the structure constants of the isomorphic Lie algebras
$s(M_n) \sim i(M_n)$.
\end{theor}
\begin{demo}
The proof is based on the identities given in \cite{ES}.
\end{demo}\\
The natural consequence is
\begin{cor}\label{adjX}
The operators $U_{\xi}\in spin[S(M_n)]$ transform the basis generators $X_a$
according to the adjoint representation of $S(M_n)$,
\[
U_{\xi}X_aU^{-1}_{\xi}=Adj(\xi)_{a b}X_b\,,
\]
defined as
\[
Adj(\xi)=e^{i\xi^a adj(X_a)}\,, \quad adj(X_a)_{bc}=-ic_{abc}\,,
\]
where $adj(X_a)$ are the basis generators of the adjoint representation of
$s(M_n)$.
\end{cor}
\begin{demo} This is a general result of the group representation theory
 \cite{BR}. We note that here the phase factors are chosen such that
the commutators
\[
[adj(X_a),\, adj(X_b)]=ic_{abc} \, adj(X_c)
\]
keep  the form (\ref{comX}).
\end{demo}\\
Whenever one can define a convenient relativistic scalar product in
order to get only  self-adjoint generators, the representation $spin[S(M_n)]$ 
is {\em unitary}. In this case
one can define quantum modes correctly, using the set of commuting operators
formed by the Casimir operators of $spin[s(M_n)]$, the generators of its Cartan
subalgebra and the Dirac operator, $D$.

\section{Special geometries}

Apart from the standard Dirac operator, other operators of the same
type may be defined with the help of suitable geometric objects as,
for example, the K-Y tensors with some supplemental properties,
known as complex structures or unit roots \cite{CV7}. However, such
objects arise only in geometries with particular features.

\subsection{K\" ahlerian geometries}

A special type of geometries with many possible applications in physics is
represented by the K\" ahlerian manifolds. In general, these are manifolds
$M_n$ of even dimension, $n=2k$, equipped with special geometric objects
called {\em  complex structures}. The complex structures
are  particular automorphisms $h: {\cal T}(M_{n}) \to {\cal T}(M_{n})$, of
the tangent fiber bundle ${\cal T}(M_{n})$, which are {\em covariantly
constant} and satisfy the algebraic property of a {\em complex unit}.
This means that the matrix of $h$ in a given basis, denoted by
$\left<h\right>$, must satisfy the condition
\begin{equation}\label{hh1}
\left<h\right>^2=-1_{n}
\end{equation}
(where $1_{n}$ is the $n\times n$ identity matrix). The matrix
$\left<h\right>$ in local frames is a pseudo-orthogonal
point-dependent transformation of the gauge group
$G(\eta)=SO(n_+,n_-)$ of $M_n$. With the help of the complex
structure $h$ one gives the following definition \cite{LM,GM}:
\begin{defin}
A Riemannian metric $g$ on $M_n$ is said K\" ahlerian if $h$ is
pointwise orthogonal, i.e., $g(hX,hY)=g(X,Y)$ for all $X,Y\in
{\cal T}_x(M_{n})$ at all points $x$.
\end{defin}
In natural frames  the matrix $\left<h\right>$ has the elements
$h^{\mu\cdot}_{\cdot\nu}$, defining  a skew-symmetric second rank
tensor with real-valued covariant components $h_{\mu\nu}=-h_{\nu\mu}$
which obey the condition $g_{\mu\nu}h^{\mu\,\cdot}_{\cdot\,\alpha}
h^{\nu\,\cdot}_{\cdot\,\beta}=g_{\alpha\beta}$ resulted from
Eq. (\ref{hh1}). The tensor $h$ gives rise to the
symplectic form $\tilde\omega=\frac{1}{2}h_{\nu\mu}dx^{\nu}\land dx^{\mu}$
(which is closed and non-degenerate).
However, alternative definitions can be formulated starting with both, $g$ and
$\tilde\omega$, which have to satisfy the K\" ahler relation
$\tilde\omega(X,Y)=g(X,hY)$ \cite{GM}.

A {\em hypercomplex structure} on $M_n$ is an ordered triplet ${\bf h} =
\{h^1, h^2, h^3\}$ of complex structures on $M_n$ satisfying
\begin{equation}\label{algf0}
{{\textstyle  \left<h^i\right>\,\left<h^j\right>}=-\delta_{ij} 1_n} +
\varepsilon_{ijk} {\textstyle
\left< \right.} h^k {\left. \textstyle \right>}\,,
\quad i,j,k...=1,2,3\,.
\end{equation}
In Lie algebraic terms, the matrices ${1 \over 2} \left<h^j\right>$ realize
the $su(2)$ algebra.
\begin{defin}
A hyper-K\" ahler manifold is a manifold whose Riemannian metric is
K\"ahlerian with respect to each different complex structures
$h^1, h^2$ and $h^3$.
\end{defin}

\subsection{Isolated unit roots}

The complex structures are real-valued K-Y tensors since the
classical geometric objects are in general real fields. However, the
theories with spin could involve even complex-valued K-Y tensors
generating new Dirac-type operators. Based on this argument, we
considered complex-valued complex structures called unit roots
\cite{CV7}.
\begin{defin}
The non singular real or complex-valued K-Y tensor $f$ of rank 2 defined on
$M_n$ which satisfies
\begin{equation}\label{ffg}
f^{\mu\cdot}_{\cdot\alpha}f_{\mu\beta}=g_{\alpha\beta}
\end{equation}
is called an unit root of the metric tensor $g$ of $M_n$, or simply an unit
root of $M_n$.
\end{defin}
One can show that the unit roots are covariantly constant K-Y tensors which
satisfy the algebraic condition $\left<f\right>^2=-1_n$, equivalent to
Eq. (\ref{ffg}).

It is known that the unit roots are allowed only by manifolds $M_n$
with an even number of dimensions, $n=2k$. Our previous results show
that the matrices of the unit roots are unimodular (i.e., ${\rm det}
\left<f\right>=1$)  and completely reducible in $2\times 2$ diagonal
blocks \cite{CV7}. Hereby it results
\begin{theor}
The unit roots are usual complex structures with
real matrices only when $n_+$ and $n_-$ are even numbers. In the case
of odd values of $n_+$ and $n_-$ the unit roots have complex-valued components.
\end{theor}
\begin{demo}
The proof is given in Ref. \cite{CV7}.
\end{demo}\\
Thus we understand that the unit roots are defined in a similar way
as the complex structures with the difference that the unit roots
are automorphisms of the {\em complexified} tangent bundle, $f:
{\cal T}(M_{n})\otimes{\Bbb C} \to {\cal T}(M_{n})\otimes {\Bbb C}$.
Therefore, when $f$ has complex-valued components,  the matrix
$\left<f\right>$ is a transformation of the complexified group
$G_{c}(\eta)$.

Each unit root $f$ can be seen as the basis of the one-dimensional
linear real space $L_f=\{\rho|\rho=\alpha f,\, \alpha \in {\Bbb
R}\}$ whose elements apart from ${0}$ are called roots since
$\left<\rho\right>^2=-\alpha^2 1_n$. The quantity $|\alpha|$ play
here the role of the {\em norm} of the root $\rho=\alpha f$. When
the unit root $f$ is complex-valued, there exists the {\em adjoint}
root $f^*$ which differs from $f$ and satisfies $[\left<f\right>^*,
\left<f\right>]=0$. The unit root $f^*$ is the basis of the linear
space of roots $L_{f^*}$, defined in the same manner as $L_f$. As
observed in Ref.\cite{CV7} the spaces $L_f$ and $L_{f^*}$ can not be
embedded in a larger linear structure.

\subsection{Triplets of unit roots}

The next step is to look for manifolds allowing families of unit
roots ${\bf f}=\{f^i|i=1,2,...N_f\}$ having supplementary properties
which should guarantee that: (I) the linear space $L_{\bf
f}=\{\rho|\rho=\rho_i f^i,\, \rho_i\in {\Bbb R}\}$ is isomorphic
with a real Lie algebra, and (II) each element of $L_{\bf f}-{0}$ is
a root (of arbitrary norm). In these circumstances we have
demonstrated the following theorem
 \begin{theor}\label{bumbum}
The unique type of family of unit roots with $N_f>1$ having the
properties (I) and (II) are the triplets ${\bf f}=\{ f^1,f^2,f^3\}$ which
satisfy
\begin{equation}\label{algf}
{{\textstyle  \left<f^i\right>\,\left<f^j\right>}=-\delta_{ij} 1_n} +
\varepsilon_{ijk} {\textstyle
\left< \right.} f^k {\left. \textstyle \right>}\,,
\quad i,j,k...=1,2,3\,.
\end{equation}
\end{theor}
\begin{demo}
We delegate the proof to Ref. \cite{CV7}. \end{demo}\\
Hereby it results that the matrices  $\left<f^i\right>$ and $1_n $ generate a
matrix representation of the quaternion algebra ${\Bbb H}$.
If the unit roots $f^i$ have only real-valued components
we recover the  hypercomplex structures that obey Eq. (\ref{algf0}) and define
the hyper-K\" ahler geometry. An example of hyper-K\" ahler manifold
is the Euclidean Taub-NUT space which is equipped with only one family of
real unit roots  \cite{CV6, CV7}.

The manifolds with pseudo-Euclidean metric $\eta$ of odd signature
have only pairs of {\em adjoint} triplets, ${\bf f}$ and ${\bf
f}^*$, the last one being formed by the adjoints of the unit roots
of ${\bf f}$. The spaces $L_{\bf f}$ and $L_{{\bf f}^*}$ are
isomorphic between themselves (as real vector spaces) and all the
properties of ${\bf f}^*$ can be deduced from those of ${\bf f}$
using complex conjugation. Moreover, we must specify that the set
$L_{\bf f}\bigcup L_{{\bf f}^*}$ is no more a linear space since the
linear operations among the elements of $L_{\bf f}$ and $L_{{\bf
f}^*}$ are not allowed \cite{CV7}. An  example is the Minkowski
spacetime which has a pair of conjugated triplets of complex-valued
unit roots \cite{K2}.

The mentioned  examples of manifolds possessing triplets with the properties
(\ref{algf}) are of dimension four. The results we know  indicate that similar
properties may occur for other manifolds of dimension $n=4k,\, k=1,2,3,...$
where we expect to find many such triplets  \cite{GaMo}. The main geometric
feature of all these manifolds is given by
\begin{theor}\label{ricci}
If a manifold $M_n$ allows a triplet of unit roots then this must be
Ricci flat (having $R_{\mu\nu}=0$).
\end{theor}
\begin{demo}
As in the case of the hyper-K\" ahler manifolds,
we start with the identity
$0=f_{\mu\nu;\alpha;\beta} -f_{\mu\nu;\beta;\alpha}=
f_{\mu\sigma} R^{\sigma}_{\cdot\,\nu\alpha\beta}+
f_{\sigma\nu} R^{\sigma}_{\cdot\,\mu\alpha\beta}$,
giving
\begin{equation}
R_{\mu\nu\alpha\beta}f^{\mu\,\cdot}_{\cdot\,\sigma} f^{\nu\,\cdot}_{\cdot\,
\tau}
=R_{\sigma\tau\alpha\beta}\,,
\end{equation}
and calculate $R_{\mu\nu\alpha\beta}f^{1\,\alpha\beta}=
R_{\mu\nu\sigma\beta}f^{3\, \sigma\,\cdot}_{~\,\,\cdot\,\alpha}
(\left<f^3\right>\left<f^{1}\right>)^{\alpha\beta}=
R_{\mu\nu\sigma\beta}f^{3\, \sigma\,\cdot}_{~\,\,\cdot\,\alpha}
f^{2\,\alpha\beta}=$\\
$-R_{\mu\nu\alpha\beta}f^{1\,\alpha\beta}=0\,.$
Then, permutating the first  three indices of $R$ we find the identity
\begin{equation}\label{2Rf0}
2R_{\mu\alpha\nu\beta}f^{1\,\alpha\beta}=
R_{\mu\nu\alpha\beta}f^{1\,\alpha\beta}=0\,.
\end{equation}
Finally, writing
$R_{\mu\nu}=
R_{\mu\alpha\nu\beta}f^{1\,\alpha\,\cdot}_{~\,\cdot\,\tau} f^{1\,\beta\tau}=
-R_{\mu\alpha\sigma\beta}
f^{1\,\sigma\,\cdot}_{~\,\cdot\,\nu} f^{1\,\alpha\beta} =0$, 
we draw the conclusion that the manifold is Ricci flat. The
same procedure holds for $f^2$ or $f^3$ leading  to identities similar
to (\ref{2Rf0}).
\end{demo}\\
Note that the manifolds possessing only single unit roots
(as the K\" ahler ones) are not forced to be Ricci flat.

The passing from the complex structures to unit roots has to be
productive from the Dirac theory where the complex-valued K-Y
tensors could be involved in the theory of the Dirac-type operators.

\section{Dirac-type operators}

First of all, the unit roots are complex-valued covariantly constant
K-Y tensors. It is known that any second rank K-Y tensor $f$ (having
the components $f_{\mu\nu}=-f_{\nu\mu}$ which satisfy the Killing
equation, $f_{\mu\nu;\sigma}+f_{\mu\sigma;\nu}=0$) gives rise to a
specific first-order operator of the Dirac type acting in the space
$\Psi$.
\begin{defin}
The operators
\begin{equation}\label{df}
D_{f} = i\gamma^\mu \left(f_{\mu\,\cdot}^{\cdot\,\nu}\nabla_\nu
- \textstyle{1\over 6} f_{\mu\nu;\rho}\gamma^\nu \gamma^\rho \right)\,,
\end{equation}
constructed with the help of second rank K-Y tensors, $f$, are
called  Dirac-type operators.
\end{defin}
These are non-standard Dirac operators which have the remarkable
property to anticommute with the standard Dirac operator,
$\{D_{f},\,D\}=0$ \cite{CML}. These operators can be involved in new
types of (super)symmetries. Moreover, new interesting superalgebras
of Dirac-type operators can be obtained when the second-order K-Y
tensors we use are unit roots.

\subsection{Operators produced by isolated unit roots}

The Dirac-type operator generated by an unit roots $f$ has the simpler form
\begin{equation}\label{Dirf}
D_f=if_{\mu\,\cdot}^{\cdot\,\nu}\gamma^{\mu}\nabla_{\nu}\,,
\end{equation}
since the last term of Eq. (\ref{df}) vanishes in the virtue of the
fact that $f$ is now covariantly constant. The operators of this kind
have  an important property formulated in Ref.  \cite{K1}.
\begin{theor}\label{DtyD}
The Dirac-type operator $D_f$ produced by the K-Y tensor $f$
satisfies the  condition
\begin{equation}\label{D2D2}
(D_{f})^2=D^2 \,.
\end{equation}
if and only if $f$ is an unit root.
\end{theor}
\begin{demo}
Arguments are given in  \cite{K1} showing that Eq.  (\ref{D2D2})
holds only for unit roots.
Moreover, we note that for any unit root $f$  the square of
the Dirac-type operator (\ref{Dirf}) has to be calculated exploiting  the 
identity
$0=f_{\mu\nu;\alpha;\beta} -f_{\mu\nu;\beta;\alpha}=
f_{\mu\sigma} R^{\sigma}_{\cdot\,\nu\alpha\beta}+
f_{\sigma\nu} R^{\sigma}_{\cdot\,\mu\alpha\beta}$,
which gives
\begin{equation}\label{D2D}
R_{\mu\nu\alpha\beta}f^{\mu\,\cdot}_{\cdot\,\sigma} f^{\nu\,\cdot}_{\cdot\,
\tau}=R_{\sigma\tau\alpha\beta}
\end{equation}
and leads to Eq.  (\ref{D2D2}).
\end{demo}\\
Thus we  conclude that the condition (\ref{D2D2}) holds in any geometry of 
dimension $n=2k$
allowing unit roots (or complex structures).

Another interesting operator related to $f$ can be defined as follows.
\begin{defin}
Given the unit root $f$, the matrix
\begin{equation}\label{SfS}
\Sigma_{f}=\frac{1}{2}f_{\mu\nu}S^{\mu\nu}
\end{equation}
is the spin-like operator associated to $f$.
\end{defin}
This is a matrix that acts on the space of spinors $\Psi$ and, therefore,
can be interpreted as a generator of the spinor representation
$spin[{\bf G}(\eta)]$. From Eqs. (\ref{Nabg}) and (\ref{ffg}) one
can verify that this generator is covariantly constant in the sense that
$\nabla_{\nu}(\Sigma_f\psi)=\Sigma_f \nabla_{\nu} \psi$.
Hereby we find that the Dirac-type operator (\ref{Dirf}) can be written as
\begin{equation}\label{DDS}
D_f=i\left[D,\, \Sigma_f\right] \,,
\end{equation}
where $D$ is the standard Dirac operator defined by Eq.  (\ref{Dirac}).
In addition, we obtain the useful relations  
$[\Sigma_f,\, D^2]=[\Sigma_f,\,(D_{f})^2]=0$\,.

When there is a complex-valued unit root $f$ then $f^*\not
=f$ and the corresponding spin-like operators are related to each
other as $\Sigma_{f^*}=\overline{\Sigma}_f$ and, therefore, $\Sigma_{f^*}$
commutes with $D^2$. The Dirac-type operators
$D_f$ and $D_{f^*}=i[D,\Sigma_{f^*}]=\overline{D}_f$ satisfy
$(D_f)^2=(D_{f^*})^2=D^2$ and
\[
\left\{D_f,D\right\}=0\,,\quad \left\{D_{f^*},D\right\}=0\,.
\]

These properties suggest us to define ${\cal N}=2$ superalgebras of
Dirac operators.
\begin{defin}
Given an isolated unit root $f$, we say that the set ${\bf
  d}_f=\{D(\lambda)|D(\lambda)=\lambda_0 D+\lambda_1
  D_f\,;\lambda_0,\lambda_1\in {\Bbb R}\}$ represent the ${\cal N}=2$
 real D-superalgebra generated by the unit root $f$. The complex
D-superalgebra with the same generators, $D$ and $D_f$, but having
complex-valued coefficients will be denoted by $({\bf d}_f)_c$.
\end{defin}
When $f^*\not = f$ the D-superalgebra ${\bf d}_{f^*}$ differs from ${\bf
  d}_f$ and, in general, these can not be embedded in a larger superalgebra.

\subsection{Operators produced by triplets of unit roots}

In the case of manifolds allowing triplets  of unit roots, ${\bf
f}=\{f^1,f^2,f^3\}$, satisfying Eq. (\ref{algf}), one can construct
spin-like and Dirac-type operators for any unit root $f^i$. Then it
is convenient to introduce the simpler notations
$\Sigma_i\equiv\Sigma_{f^i}$ and $D^i\equiv
D_{f^i}=i[D,\,\Sigma^i]$, $i=1,2,3$. These operators have the same
properties as those produced by isolated unit roots. In addition,
from Eqs. (\ref{algf}) we deduce that
\begin{equation}\label{SSffS}
\left[\Sigma^i, \Sigma^j\right]= \textstyle{\frac{i}{2}
\left[\,\left<f^i\right>,\,\left<f^j\right>\, \right]_{\mu\nu}}
S^{\mu\nu}= 2i\varepsilon_{ijk}\Sigma^k   \,.
\end{equation}
Taking into account that $\Sigma^i$ are covariantly constant we find
other important relations,
\[
\left[D^i, \Sigma^j\right]=i\delta_{ij}D+i\varepsilon_{ijk}D^k\,.
\]
When we work with triplets of
complex-valued unit roots, we must  consider the operators generated
by the adjoint triplet ${\bf f}^*$ whose properties have to be
obtained simply using the Dirac conjugation as in the previous case
of isolated unit roots. Thus it is not difficult to show that
$\Sigma_{(f^i)^*}=\overline{\Sigma}^i$ and
$D_{(f^i)^*}=\overline{D}^{i}$.

The triplets generates larger superalgebras whose algebraic structure is 
provided by
\begin{theor}
If a triplet ${\bf f}$ accomplishes  Eqs.  (\ref{algf})
then the corresponding Dirac-type operators satisfy
\begin{equation}\label{4sup}
\left\{D^i,\,D^j\right\}=2\delta_{ij} D^2\,,\quad
\left\{D^i,\,D\right\}=0\,.
\end{equation}
\end{theor}
\begin{demo}
If $i=j$ we take over the result of Theorem \ref{DtyD}. For
$i\not=j$ we take into account that $M_n$ is Ricci flat  finding
that $D^i$ and $D^j$ anticommute. The second relation was
demonstrated earlier for any unit root.
\end{demo}\\
Thus it is obvious that the operators $D$ and $D^i$ ($i=1,2,3$) form a
basis of a four-dimensional  superalgebra of Dirac operators.
\begin{defin}
The set ${\bf
  d}_{\bf f}=\{D(\lambda)|D(\lambda)=\lambda_0 D+\lambda_i
  D^i\,;\lambda_0,\lambda_i\in {\Bbb R}\}$ represent the ${\cal N}=4$
 real D-superalgebra generated by the triplet ${\bf f}$. The complex
D-superalgebra $({\bf d}_{\bf f})_c$ has the same basis but
  complex-valued coefficients, $\lambda_0, \lambda_i \in {\Bbb C}$.
\end{defin}
If the adjoint triplet  ${\bf f}^*$ differs from ${\bf f}$ then the
(real or complex) D-super\-al\-ge\-bras generated by these triplets are
different to each other and can not be seen as subspaces of a larger
linear structure. The relation between these D-superalgebras is
quite complicated and one must clear it up for each particular case
separately.

\section{Symmetries generated by unit roots}

The D-superalgebras have the interesting property  to present a new
type of symmetries due only to the existence of the unit roots
\cite{CV7}. These give the spin-like operators which  generate
transformations relating the Dirac-type operators and the standard
Dirac one among themselves. These symmetries mixed with the
transformations produced by isometries give the complete set of
automorphisms of the above defined D-superalgebras.

\subsection{Symmetries generated by isolated unit roots}

In the simpler case of manifolds allowing only isolated
unit roots $f$ we have to study the continuous symmetry generated by
the Lie algebra $L_f$ which is able to relate $D$ and $D_f$ to each other.
Let us  observe that the roots $\rho=\alpha f\in L_f$  define the sections
$A_{\rho} : M_n \to {\bf G}_c(\eta)$ of the complexified spinor fiber bundle 
such that
$A_{\rho}(x)=[\rho(x)]\in {\bf G}_c(\eta)$.
\begin{defin}
We say that
${G}_{f}=\{(A_{\rho},id)\,|\,\rho=\alpha f,\, \alpha \in {\Bbb R}\}
\subset [{\cal G}_c(M_n)]$ is the one-parameter Lie group associated to
the unit root $f$.
\end{defin}
The spinor representation of this group, $spin(G_f)$,
is formed by all operators $U_{\rho}\in spin{\cal G}(M_n)$ with
$\rho=\alpha f$ whose  transformation matrices have the form
\begin{equation}\label{Taf}
T(\alpha f) =e^{-i\alpha \Sigma_f} \in spin[{\bf G}_c(\eta)]\,,
\end{equation}
depending on the group parameter $\alpha \in{\Bbb R}$.  Since the matrices 
(\ref{Taf}) are
just those defined by Eq.
(\ref{TeS}) where we replace $\omega$ by the roots $\rho=\alpha f\in L_f$,
their action on the point-dependent Dirac matrices results from Eq.
(\ref{TgT}) to be,
\begin{equation}\label{TgT1}
[T(\alpha f)]^{-1}\gamma^{\mu}T(\alpha f)=
\Lambda^{\mu\,\cdot}_{\cdot\,\nu}(\alpha f)\gamma^{\nu}\,,
\end{equation}
where
$\Lambda^{\mu\,\cdot}_{\cdot\,\nu}=
e^{\mu}_{\hat\alpha}\Lambda^{\hat\alpha\,\cdot}_{\cdot\,\hat\beta}
\hat e^{\hat\beta}_{\nu}$ are matrix elements with natural indices of the
matrix
\begin{equation}\label{Lamaf}
\Lambda(\alpha f)=e^{\alpha \left<f\right>}=1_n\cos\alpha+
\left<f\right>\sin\alpha \,,
\end{equation}
calculated according to Eqs. (\ref{Lam})  and (\ref{ffg}). We note that this
is a matrix representation of the usual {\em Euler formula} of the complex
numbers. Now it is obvious that in local frames
$\left<f\right>=\Lambda(\frac{\pi}{2}f)\in vect[{\bf G}_c(\eta)]$, as mentioned
above.
\begin{theor}\label{TDT}
The operators ${U}_{\rho}\in spin(G_f)$, with $\rho=\alpha f$,
have the following action in the linear space spanned by the operators
$D$ and $D_f$:
\begin{eqnarray}
&&~{U}_{\rho}D({U}_{\rho})^{-1}
=T(\alpha f) D [T(\alpha f)]^{-1}=D\cos\alpha+D_f\sin\alpha\,,\label{T1}\\
&&{U}_{\rho}D_f({U}_{\rho})^{-1}
=T(\alpha f) D_f [T(\alpha f)]^{-1}=-D\sin\alpha+D_f\cos\alpha\,.
\label{T2}
\end{eqnarray}
\end{theor}
\begin{demo}
From Eq.  (\ref{Lamaf}) we obtain the matrix elements
$\Lambda^{\mu\,\cdot}_{\cdot\,\nu}(\alpha f)=
\cos\alpha\, \delta^{\mu}_{\nu}+\sin\alpha\,
f^{\mu\,\cdot}_{\cdot\,\nu}$
which lead to the above result since $\Sigma_f$ as well as $T(\alpha f)$
are covariantly constant.
\end{demo}\\
From this theorem it results that $\alpha\in [0,2\pi]$ and, consequently,
the group $G_f\sim U(1)$ is {\em compact}. Therefore, it must be
a subgroup of the maximal compact subgroup of ${\bf G}_c(\eta)$.
Note that the transformations (\ref{T1}) and (\ref{T2})
leave the operator $D^2=(D_f)^2$ invariant  because this commutes
with the spin-like operator $\Sigma_f$ which generates these transformations.

In general, when $f$ has complex components (and $f^*\not=f$) then
$L_{f^*}\sim so(2)$ is a different linear space representing the Lie
algebra of $vect({G}_{f^*})$. These two Lie algebras are complex
conjugated to each other and, therefore, remain isomorphic (as real
or complex algebras).  The relation among the transformation
matrices of $spin({G}_{f})$ and $spin({G}_{f^*})$ is
$\overline{T}(\alpha f)=T(-\alpha f^*)=[T(\alpha f^*)]^{-1}$ which
means that when $f^*\not=f$ the representation $spin(G_f)$ is no
longer unitary in the sense of the generalized Dirac adjoint.
However, if $f^*=f$  is a complex structure then $M_n$ is an usual
K\" ahler manifold and the representation $spin(G_f)$ is unitary.

The conclusion is that a unit root $f$ gives rise  simultaneously to a
Dirac-type operator $D_f$ which satisfies Eq.  (\ref{D2D2}) and the
one-parameter  Lie group $G_f$ which relates $D$ and $D_f$ to each other.
This general result can be extended for the triplets too.

\subsection{Symmetries generated by triplets of unit roots}

In manifolds with  triplets ${\bf f}=\{f^1,f^2,f^3\}$ we shall find
more complicated symmetries whose properties are provided by
Eqs. (\ref{algf}) combined with the previous results (\ref{SSffS})-(\ref{LX}).
\begin{defin}
We say that $G_{\bf f}=\{(A_{\rho},id)\,|\,\rho 
=\rho_if^i\in L_{\bf f}\}\sim SU(2)
\subset {\bf G}_c(\eta)$ is the Lie group associated to the triplet ${\bf f}$.
\end{defin}
The spinor and the vector representations of this group are determined by the
representations of its Lie algebra, $g_{\bf f}$, resulted from
Theorem \ref{bumbum}.
\begin{theor}
The basis-generators of $vect(g_{\bf f})=L_{\bf f}$ are
$J_i=\frac{i}{2} \left<f^i\right>$ while the
basis-generators of the algebra $spin(g_{\bf f})\sim su(2)$ read
$\hat s_i=\frac{1}{2}\Sigma^i$, $(i=1,2,3)$.
\end{theor}
\begin{demo}
From Eqs.  (\ref{algf}) and (\ref{SSffS})  we deduce that these
generators satisfy the standard  $su(2)$ commutation rules. We note
that they are Hermitian only in the K\" ahlerian case when ${\bf
f}^*={\bf f}$.
\end{demo}\\
The group $vect(G_{\bf f})=\{\Lambda(A_{\rho})\,|\,\rho\in L_{\bf f},\,
\|\rho\|\le 2\pi\}$ of the vector representation is the compact group formed by the
matrices
\begin{equation}\label{LX}
\Lambda(\rho)=\textstyle{e^{\rho_i \left<f^i\right>}=1_n\cos\|\rho\|+
\nu_i \left<f^i\right>}\sin\|\rho\| \,,
\end{equation}
where $\|\rho\|=\sqrt{\rho_i\rho_i}$  and $\nu_i =\rho_i/\|\rho\|$ for
any $\rho\in L_{\bf f}\sim su(2)\sim so(3)$.
The action of the operators of the spinor representation
$U_{\rho}\in spin(G_{\bf f})$ is defined by the transformation
matrices
\begin{equation}\label{TxiS}
T(\rho)=e^{-i\rho_i \Sigma^i} =e^{-2i\rho_i \hat s_i}\,,
\quad \rho=\rho_i f^i\in L_{\bf f}\,.
\end{equation}
These have to be calculated directly from Eq.  (\ref{Taf}) replacing
$\alpha=\pm\|\rho\|$ and $f=\pm \rho/ \|\rho\|$. Then the
transformations (\ref{TgT1}) can be expressed in terms of parameters
$\rho_i$ using the matrices $\Lambda({\rho})$.

Hereby we observe that $\rho_i$ are nothing other than
the analogous of the well-known Caley-Klein parameters but ranging in a larger
spherical domain (where $\|\rho\|\le 2\pi$) such that they cover twice the
group $G_{\bf f}\sim SU(2)$, as we can convince ourselves calculating
\begin{equation}\label{LfLf}
\textstyle{\Lambda(\rho)\left<f^i\right>\Lambda^T(\rho)
={\frak R}_{ij}(2\vec{\rho})
\left<f^j\right>}\,, \quad \forall\, \Lambda_{\rho}\in vect(G_{\bf f})\,,
\end{equation}
where ${\frak R}(2\vec{\rho})\in O(3)$ is the rotation of the Caley-Klein
parameters $2\rho_i$. These arguments lead to the conclusion  that
$vect(G_{\bf f})\sim SU(2)$  \cite{CV6}.
On the other hand, since the rotations (\ref{LfLf}) change the basis of
$L_{\bf f}$ leaving Eqs.  (\ref{algf}) invariant, we understand that these
form the group ${\rm Aut}(L_{\bf f})$, of the automorphisms of the Lie algebra
$L_{\bf f}$ considered as a real algebra.

The most important result here is that the operator $D(\vec{\nu})=\nu_i D^i$ defined
by the unit vector  $\vec{\nu}$ (with $\vec{\nu}^2=1$)
can be related to $D$ through the transformations
(\ref{T1}) and (\ref{T2}) of the one-parameter subgroup ${G}_{f_{\vec\nu}}$
associated to the unit root $f_{\vec{\nu}}=\nu_i f^i$.
This is a subgroup of the group $G_{\bf f}$ which embeds
all the subgroups defined by  the unit roots $f_{\vec{\nu}}$.
\begin{theor}
The operators $U_{\rho}\in spin(G_{\bf f})$, with $\rho=\rho_i f^i=\|\rho\| 
\nu_i f^i$,
transform the Dirac operators  $D$, $D^i$ ($i=1,2,3$) as
\begin{eqnarray}
{U}_{\rho}D[{U}_{\rho}]^{-1}&=&
T(\rho)D[T(\rho)]^{-1}=D\cos \|\rho\|+\nu_i D^i\sin\|\rho\|\,,\label{TDT10}\\
{U}_{\rho}D^i[{U}_{\rho}]^{-1}&=&
T(\rho)D^i[T(\rho)]^{-1}\nonumber\\
&=&D^i\cos \|\rho\|-
(\nu_i D+\varepsilon_{ijk} \nu_j D^k)\sin\|\rho\|\,. \label{TDT20}
\end{eqnarray}
\end{theor}
\begin{demo}
We consider the result of Theorem (\ref{TDT}) for each one-parameter
subgroup of $G_{\bf f}$ generated by the unit root
$f_{\vec{\nu}}=\nu_if^i$.
\end{demo}

In the non-K\" alerian manifolds equipped with pairs of adjoint
triplets ${\bf f} \not = {\bf f}^*$,  the corresponding Dirac-type
operators $D^i$ and $\overline{D}^{i}$ are different to each other.
In other respects, the generators $\Sigma^i$ are not Hermitian and
this forces us to operate with non-unitary representations of the
group $G_{\bf f}\sim SU(2)$. In these conditions it may be necessary
to extend the symmetry groups considering the complexified group
$(G_{\bf f})_c\sim SL(2,{\Bbb C})$ of $G_{\bf f}$ having the same
transformations (\ref{TDT10}) and (\ref{TDT20}) but with
complex-valued parameters $\rho_i\in {\Bbb C}$ \cite{CV7}. In  this
case one doubles the number of generators of the spinor or vector
representations. More precisely,  the basis generators of the spinor
representation of the complexified algebra  $spin(g_{\bf f})_c$ are
$\hat s_i=\frac{1}{2}\Sigma^i$ and $\hat
r_i=(\pm)\frac{i}{2}\Sigma^i$ while the corresponding vector
representation, $vect(g_{\bf f})_c$, is generated by
$J_i=\frac{i}{2}\left<f^i\right>$ and
$K_i=\pm\frac{1}{2}\left<f^i\right>$. Obviously, according to Eqs.
(\ref{algf}) and (\ref{SSffS}), these generators satisfy the
standard $sl(2,{\Bbb C})$ commutation rules,
\begin{equation}\label{sss}
[\hat s_i, \hat s_j]=i\varepsilon_{ijk}\hat s_k\,,\quad
[\hat s_i, \hat r_j]=i\varepsilon_{ijk}\hat r_k\,,\quad
[\hat r_i, \hat r_j]=-i\varepsilon_{ijk}\hat s_k\,,
\end{equation}
and similarly for $J_i$ and $K_i$. The spinor and vector
representations of the group $(G_{{\bf f}^*})_c$ are generated by 
$\overline{\hat s}_i$
and $\overline{\hat r}_i$ and respectively $J_i^*$ and $K_i^*$.

\section{The effect of isometries}

Beside the types of continuous  symmetries we have studied, the
isometries could have some non-trivial effects transforming the unit
roots as well as  the Dirac-type operators.

\subsection{The invariance of isolated unit roots}

Let us start with manifolds $M_n$ with isolated unit roots
$f$ generating  ${\cal N}=2$ {\em real} superalgebras,
${\bf d}_{f}$. The basis $(D,D_f)$
can be changed through the transformations (\ref{T1}) and (\ref{T2})
that preserve the anticommutation relations.
Whenever the manifold  has a non-trivial isometry group $I(M_n)$ then an 
arbitrary
isometry $x\to x'=\phi_{\xi}(x)$  transforms $f$ as a second rank
tensor,
\begin{equation}\label{ffprim}
f_{\mu\nu}(x)\to f'_{\mu\nu}(x')
\frac{\partial x^{\prime\,\mu}}{\partial x_\alpha}
\frac{\partial x^{\prime\,\nu}}{\partial x_\beta}=
f_{\alpha\beta}(x)\,.
\end{equation}
Since there is only one $f$ we are forced to put $f'=f$ which means
that this remains {\em invariant} under isometries.
\begin{theor}
${\rm Aut}({\bf d}_f)=spin( G_{\bf f})\sim SO(2)$ is the group  of 
automorphisms of the
D-superalgebra ${\bf d}_f$.
\end{theor}
\begin{demo}
This group must be Abelian since ${\cal N}=2$. Thus there are no
transformations different from (\ref{T1}) and (\ref{T2}). This is in
accordance with the invariance of $f$ under isometries.
\end{demo}\\
A consequence is given by
\begin{cor}\label{invf}
If a manifold $M_n$ with the external symmetry group $S(M_n)$ has a
single unit root, $f$, then every generator $X\in spin[s(M_n)]$
commutes  with $D_f$.
\end{cor}
\begin{demo}
We calculate first the derivatives with respect to $\xi^a$ of Eq. 
(\ref{ffprim})
for $f'=f$ and $\xi=0$. Then, taking into account that $f$ is covariantly
constant we can write $f_{\alpha\lambda}k_{~;\beta}^{\lambda}=
f_{\beta\lambda}k_{~;\alpha}^{\lambda}$ for each Killing vector field $k$
defined by Eq.  (\ref{ka}). This identity yields
\[
[X,\, \Sigma_f]=0\,,\quad [X,\,D_{f}]=0\,, \quad \forall\, X\in spin[s(M_n)]\,,
\]
which means that the operators $\Sigma_f$ and $D_f$ are invariant under
isometries.
\end{demo}

\subsection{The transformation of triplets under isometries}

The case of the hyper-K\" ahler manifolds is  more complicated since
a triplet ${\bf f}$ gives rise to Dirac-type operators
$D^i=\overline{D}^i$ which anticommute with $D$ and present the
non-Abelian continuous symmetry discussed in the previous section.
In this case we have to study the group of automorphisms of this
D-superalgebra, ${\rm Aut}({\bf d}_{\bf f})$, and its Lie algebra,
${\rm aut}({\bf d}_{\bf f})$. The transformation matrices $T(\rho)$
commute with $D^2$, leaving Eqs. (\ref{4sup}) invariant under
transformations (\ref{TDT10}) and (\ref{TDT20}) which appear thus as
automorphisms of ${\bf d}_{\bf f}$. Consequently, the group of these
transformations, $spin(G_{\bf f})\sim SU(2)$, is a subgroup of
${\rm Aut}({\bf d}_{\bf f})$. However, we need more automorphisms in
order to complete the group ${\rm Aut}({\bf d}_{\bf f})$ with more
ordinary or invariant subgroups, isomorphic with $SU(2)$ or $O(3)$.
These supplemental automorphisms must transform the operators $D^i$
among themselves preserving their anticommutators as well as the
form of $D$. Therefore, these may be produced by the isometries of
$M_n$ since these leave the operator $D$ invariant.

In what concerns the transformation of the triplets ${\bf f}$ under isometries
we have two possibilities, either to consider that all the unit roots
$f^i\in {\bf f}$  are invariant under isometries or to
assume that the isometries transform the components of the triplet among
themselves, $f^{\prime\,i}=\hat{\frak R}_{ij}f^j$, such that Eqs.  (\ref{algf})
remain invariant. The first hypothesis is not suitable since we need more
transformations in order to fill in the group ${\rm Aut}({\bf d}_{\bf f})$ when
we do not have other sources of symmetry. Therefore, we must adopt the second
viewpoint assuming that the components of ${\bf f}$ are transformed as
\begin{equation}\label{ffprim1}
f^{j}_{\mu\nu}(x')\frac{\partial x^{\prime\,\mu}}{\partial x_\alpha}
\frac{\partial x^{\prime\,\nu}}{\partial x_\beta}=
\hat{\frak R}_{kj}(\xi,x)f^k_{\alpha\beta}(x)\,,
\end{equation}
by $3\times 3$ real or complex-valued matrices $\hat{\frak R}$ which
must be {\em orthogonal} in order to leave Eqs. (\ref{algf}) invariant.
According to these equations, the matrix elements can be put in
the equivalent forms
\begin{eqnarray}
\hat{\frak R}_{ij}(\xi,x)&= &\frac{1}{n}
f^{i\,\alpha\beta}(x)
\frac{\partial \phi_{\xi}^{\mu}(x)}{\partial x_\alpha}
\frac{\partial \phi_{\xi}^{\nu}(x)}{\partial x_\beta}
f^{j}_{\mu\nu}[\phi_{\xi}(x)] \nonumber\\
&=& \frac{1}{n}
f^{i\,\hat\alpha\hat\beta}(x)
\Lambda^{\hat\mu\,\cdot}_{\cdot\,\hat\alpha}[A_{\xi}(x)]
\Lambda^{\hat\nu\,\cdot}_{\cdot\,\hat\beta}[A_{\xi}(x)]
f^{j}_{\hat\mu\hat\nu}[\phi_{\xi}(x)]\,.\label{rotroot}
\end{eqnarray}
The last formula is suitable for calculations in local frames where we must
consider the external symmetry using the gauge transformations
$\Lambda[A_{\xi}(x)]\in vect[{\bf G}(\eta)]$ defined by Eq. (\ref{Axx}) and
associated to isometries for preserving the gauge.

The matrices $\hat{\frak R}$ might be point-dependent and depend on the
parameters $\xi^a$ of $I(M_n)$. This means that the canonical covariant
parameters $\hat\omega_{ij}=-\hat\omega_{ji}$ giving the expansion
$\hat{\frak R}_{ij}(\hat\omega)= \delta_{ij}+\hat\omega_{ij}+\cdots$
are also depending on these variables. Then, for small values of the parameters
$\xi^a$ the covariant parameters can be developed as $\hat\omega_{ij}=
\xi^a\hat c_{aij}+...$ emphasizing thus the quantities $\hat c_{aij}$
we shall see that do not depend on coordinates.
\begin{theor}
Let $M_n$ be a manifold having the triplet of unit roots
${\bf f}=\{f^1,f^2,f^3\}$ and a non-trivial isometry group
$I(M_n)$ with parameters $\xi^a$ and the corresponding Killing vectors
\[
k_a(x)=\left.\frac{\partial\phi_{\xi}(x)}{\partial \xi^a}\right|_{\xi=0}\,.
\]
Then the basis-generators (\ref{X})of the spinor representation  $spin
[s(M_n)]$ and $\Sigma^i$ satisfy
\begin{equation}\label{Xscs}
[X_a,\, \Sigma^i]=i\hat c_{aij} \Sigma^j\,, \quad a=1,2,...,N\,,
\end{equation}
where $\hat c_{aij}$ are {\em point-independent} structure constants.
\end{theor}
\begin{demo}
Deriving  Eq.  (\ref{ffprim1}) with respect to $\xi^a$ in $\xi=0$ we deduce
\begin{equation}\label{fkcf}
f^{i}_{\mu\lambda}k_{a\,;\nu}^{\lambda}-
f^{i}_{\nu\lambda}k_{a\,;\mu}^{\lambda}=\hat c_{aij}f^{j}_{\mu\nu}\,,
\end{equation}
which leads to the explicit form
\begin{equation}\label{caka}
\hat c_{aij}= -\frac{2}{n}\,\varepsilon_{ijl}\, \hat k^l_a\,,\quad
\hat k^l_a=f^{l\,\mu\nu}k_{a\, \mu;\nu}\,.
\end{equation}
Bearing in mind that  $f^i_{\mu\nu;\sigma}=0$  and using
$f^{i\,\mu\nu}k_{a\,\mu;\nu;\sigma}=
R_{\mu\sigma\nu\,\cdot}^{\,\cdot\,\cdot\,
\cdot\,\lambda}k_{a\,\lambda}f^{i\,\mu\nu}$   and Eq.  (\ref{2Rf0}),
we find  $\nabla_{\sigma}\hat k^i_a=
f^{i\,\mu\nu}k_{a\, \mu;\nu;\sigma}=0$ which means that $\partial_{\sigma}
\hat c_{aij}=0$. Finally, from Eq.  (\ref{fkcf}) we derive the commutation
rules (\ref{Xscs}).
\end{demo}\\
An important consequence is given by
\begin{cor}
The basis generators $X_a \in spin[s(M_n)]$ and the
Dirac-type operators of the ${\cal N}=4$ superalgebra ${\bf d}_{\bf f}$ obey
\[
[X_a, D^i]=i\hat c_{aij} D^j\,.
\]
\end{cor}
\begin{demo}
This formula results commuting Eq. (\ref{Xscs}) with $D$.
\end{demo}\\
We note that the structure constants $\hat c_{aij}$ take
complex values in the non-K\" ahlerian case when
${\bf f}^*\not = {\bf f}$. In other respects, the previous theorem
provides the form of the matrices ${\frak R}$.
\begin{cor}
Eq. (\ref{rotroot}) defines the point-independent $3\times 3$
matrices of the form,
\begin{equation}\label{RX}
\hat {\frak R}(\xi)=e^{i\xi^a{\frak X}_a}\,, \quad
({\frak X}_a)_{ij}=-i\hat c_{aij}\,,
\end{equation}
whose generators satisfy the commutation rules of $i(M_n)$,
\begin{equation}\label{XaXbXc}
[ {\frak X}_a, {\frak X}_b]=ic_{abc}{\frak X}_c\,.
\end{equation}
\end{cor}
\begin{demo}
These matrices are point-independent since $\hat c_{aij}$ are
structure constants. If we commute Eq.  (\ref{Xscs}) with $X_b$
using Eqs.  (\ref{comX}) and (\ref{sss}) we obtain Eq. (\ref{XaXbXc}).
\end{demo}\\
The above results lead to the following conclusion:
\begin{cor}
The group $O_{\bf f}= \{ \hat{\frak R}(\xi)\,|\, (Id,\phi_{\xi})\in
I(M_n)\}$ is a representation of the group $I(M_n)$ {\em induced}  by
the group $O_c(3)$.
\end{cor}
\begin{demo}
The matrices (\ref{RX}) are orthogonal since they are generated by
the skew-symmetric generators defined by Eqs. (\ref{RX}) and (\ref{caka}).
These matrices may have complex-valued components which means that
the representation $O_ {\bf f}$ is induced by $O_c(3)$.
\end{demo}\\
In the K\" ahlerian case, when ${\bf f}^*={\bf f}$ and $\hat
c_{aij}\in {\Bbb R}$, this representation is induced by the orthogonal
group $O(3)$.

Now we can point out how act the isometries $x\to x'=\phi_{\xi}(x)$ on the
operators $D^i$.
\begin{theor}
The Dirac-type operators $D^i$ produced by any triplet ${\bf f}$ transform
under isometries $(A_{\xi}, \phi_{\xi})\in S(M_n)$ according to the  
transformation rule
\begin{equation}\label{UDUR}
(U_{\xi}D^iU^{-1}_{\xi})[\phi_{\xi}(x)]=[T(A_{\xi})D^i {T}(A_{\xi})^{-1}](x)=
\hat{\frak R}_{ij}(\xi)D^j(x)      \,,
\end{equation}
where the action of the operator $U_{\xi}\in spin[S(M_n)]$ is defined
by Eq.  (\ref{rep}).
\end{theor}
\begin{demo}
 From Eqs.  (\ref{SfS}) and (\ref{ffprim1}) we derive
\begin{equation}\label{STTSR}
(U_{\xi}\Sigma^iU^{-1}_{\xi})[\phi_{\xi}(x)]=
[T(A_{\xi})\Sigma^i {T}(A_{\xi})^{-1}](x)=
\hat{\frak R}_{ij}(\xi)\Sigma^j(x)\,,
\end{equation}
which leads to Eq. (\ref{UDUR}) after a commutation with $D$ that is invariant
under $U_{\xi}$.
\end{demo}

\section{The automorphisms of 
${\bf d}_{\bf f}$ and $({\bf d}_{\bf f})_c$} 

Now we have a complete image of the symmetries that preserve the
anticommutation rules of the real superalgebra ${\bf d}_{\bf f}$ in a
hyper-K\" ahler manifold. These are given by the transformations
defined by Eqs.  (\ref{TDT10}), (\ref{TDT20}) and (\ref{UDUR}).
In what concerns the structure of the groups of automorphisms of the
${\cal N}=4$ D-superalgebras, the problem must be treated separately
in the K\" ahlerian and non-K\" ahlerian cases.

Let us consider first the hyper-K\" ahler manifolds.
\begin{theor}\label{gicu}
Whenever ${\bf f}^*={\bf f}$ the group $Aut({\bf d}_{\bf
f})=spin[G_{\bf f}\,\circledS\,  S(M_n)]$ is a representation of the
semi-direct product  $G_{\bf f} \,\circledS\, S(M_n)$ where $G_{\bf
f}$ is the invariant subgroup.
\end{theor}
\begin{demo}
We observe that in this case $O_{\bf f}$ is an unitary
representation of $I(M_n)$ induced by $O(3)$. This means that the
transformations (\ref{UDUR})
lead only to real linear combinations of Dirac operators such that it is
enough to study the automorphisms of the real D-superalgebra ${\bf d}_{\bf f}$.
The basis generators of the Lie algebra $aut({\bf d}_{\bf f})$ of the group
$Aut({\bf d}_{\bf f})$ are the operators $\hat s_i$ and $X_a$ that obey the
commutation relations (\ref{comX}), (\ref{sss}) and (\ref{Xscs}).
These operators form a Lie algebra since $\hat c_{aij}$ are
point-independent. In this algebra $g_{\bf f}\sim su(2)$ is an ideal and,
therefore, the corresponding $SU(2)$ subgroup is invariant. However, this 
result
can be obtained directly taking $(A_{\xi},\, \phi_{\xi})\in S(M_n)$ and
 $(A_{\rho}, id)\in G_{\bf f}$ and evaluating
$(A_{\xi},\, \phi_{\xi})* (A_{\rho},\, id)* (A_{\xi},\, \phi_{\xi})^{-1}=
([A_{\xi}\times (A_{\rho}\times A_{\xi}^{-1})]\circ \phi_{\xi}^{-1}, \,id)=
(A_{\rho'},\, id)$ where, according to (\ref{Axx}), (\ref{LX}) and
(\ref{ffprim1}), we have $\rho'=\rho_i\hat{\frak R}_{ij}(\xi)f^j$.
Consequently, $(A_{\rho'},\, id)\in G_{\bf f}$
which means that $G_{\bf f}\sim SU(2)$ is an invariant subgroup.
\end{demo}\\
An interesting restriction can be formulated as
\begin{cor}
The minimal condition that $M_n$ allows a hypercomplex structure is to
have an isometry group that includes at least one $O(3)$ subgroup.
\end{cor}
\begin{demo}
The subgroup $O(3)$ of ${\rm Aut}({\bf d}_{\bf f})$ needs at least
three generators $X_a$ satisfying the $su(2)$ algebra. Thus we conclude that
$S(M_n)$ must include one $SU(2)$ group for each different hypercomplex
structure of $M_n$.
\end{demo}\\
This restriction is known in four dimensions  where there exists only three
hyper-K\" ahler manifolds with only one hypercomplex structure and one
subgroup  $O(3)\subset I(M_4)$  \cite{GR1}. These are
given by the Atiyah-Hitchin  \cite{AH}, Taub-NUT and Eguchi-Hanson  \cite{EH}
metrics, the first one being only that does not admit more $U(1)$ isometries
\cite{GR1,BS}.

The case of the non-K\" ahlerian manifolds is more delicate since
here the matrices $\hat {\frak R}$ are complex-valued orthogonal
matrices of the complexified group $O_c(3)$. Consequently, these lead
to linear combinations of the operators $D^i$ with complex-valued
coefficients. In these circumstances we must consider the
automorphisms of the complex D-superalgebra $({\bf d}_{\bf f})_c$.
However, this requires to use the complexified group $(G_{\bf
f})_c\sim SL(2,{\Bbb C})$ instead of $G_{\bf f}$. In this way we
arrive at
\begin{theor}
If ${\bf f}^*\not ={\bf f}$ then the group $Aut({\bf d}_{\bf
f})_c=spin[(G_{\bf f})_c  \,\circledS\, S(M_n)]$ is the spin
representation of the semi-direct product  $(G_{\bf f})_c
\,\circledS\, S(M_n)$ where $(G_{\bf f})_c$ is the invariant
subgroup.
\end{theor}
\begin{demo}
The proof is the same as for Theorem (\ref{gicu}).
\end{demo}\\
Moreover, following similar arguments as in the case of the K\"
ahlerian manifolds, one can deduce the minimal condition that $M_n$
allows a pair of adjoint triplets. This requires
the group $S(M_n)$ to include at least one $SL(2, {\Bbb C})$ subgroup since we
need six generators for building the representation $O_{\bf f}\sim O(3)_c$.
The Minkowski spacetime which has a pair of adjoint triplets and $O(3,1)$
isometries is a typical example.

\section{The Minkowski spacetime}

Let us consider the Minkowski spacetime, $M$, with the metric
$g=\eta=(1,-1,-1,-1)$ and Cartesian coordinates, $x^{\mu}$
($\mu,\nu,...=0,1,2,3$) representing the time $x^0=t$ and the space
coordinates $x^i$, ($i,j,...=1,2,3$). We
choose the usual gauge of {\em inertial} frames given by
$e^{\mu}_{\nu}=\hat e^{\mu}_{\nu}=\delta^{\mu}_{\nu}$.
In this gauge we take the chiral representation of the Dirac matrices
(with off-diagonal $\gamma=\gamma^0$ \cite{TH})
where the standard Dirac operator reads
\[
D=i\gamma^{\mu}\partial_{\mu}=\left(
\begin{array}{cc}
0&i(\partial_t+\vec{\sigma}\cdot\vec{\partial})\\
i(\partial_t-\vec{\sigma}\cdot\vec{\partial})&0
\end{array}\right)=\left(
\begin{array}{cc}
0&D^{(+)}\\
D^{(-)}&0
\end{array}\right)\,,
\]
and the generators $S_{\mu\nu}$ of the spinor representation of the group
${\bf G}(\eta) = SL(2,{\Bbb C})$ have the form
\[
S^{ij}=\varepsilon_{ijk} S_k=\frac{1}{2}\varepsilon_{ijk}{\rm diag}
(\sigma_k, \sigma_k)\,,\quad
S^{i0}=\frac{i}{2}{\rm diag}(\sigma_i,-\sigma_i)\,.
\]
All these operators are self-adjoint with respect to the usual Dirac adjoint 
($\overline X=
\gamma^0 X^{+}\gamma^0$).

The isometries of $M$ are the transformations $x'=\Lambda(\omega)x
-a$ of the Poincar\' e group, $I(M)={\cal P}_{+}^{\uparrow} =
T(4)\,\circledS\, L_{+}^{\uparrow}$  \cite{W}. The group of external
symmetry $S(M)= \tilde{\cal P}^{\uparrow}_{+}\sim T(4)\,\circledS\,
SL(2,\Bbb C)$ is the universal covering group of $I(M)$. If we
denote by $\xi^{(\mu\nu)}=\omega^{\mu\nu}$ the $SL(2,\Bbb C)$
parameters and by $\xi^{(\mu)}=a^{\mu}$ those of the translation
group $T(4)$, then we obtain the standard basis generators of the
algebra $spin[s(M)]$,
\begin{eqnarray}
X_{(\mu)}&=&i\partial_{\mu} \,,\nonumber\\
X_{(\mu\nu)}&=&i(\eta_{\mu\alpha}x^{\alpha}
\partial_{\nu}- \eta_{\nu\alpha}x^{\alpha}\partial_{\mu})+
S_{\mu\nu} \,,\nonumber
\end{eqnarray}
which show us that in this gauge the Dirac field $\psi$ transforms manifestly
covariant.
In applications it is convenient to denote ${\cal P}_{\mu}=X_{(\mu)}$,
${\cal J}_i=\frac{1}{2}\varepsilon_{ijk}X_{(jk)}$ and  ${\cal K}_i=X_{(0i)}$.

\subsection{Dirac-type operators}

The Minkowski spacetime possesses a pair of adjoint triplets
${\bf f}\not = {\bf f}^* $ \cite{K2}. The
unit roots of the first triplet, ${\bf f}=\{f^1,f^2,f^3\}$, have the 
non-vanishing
complex-valued  components \cite{K2}
\begin{eqnarray}
f^1_{23}=1\,,&\quad&f^1_{01}=i\,,\nonumber\\
f^2_{31}=1\,,&\quad&f^2_{02}=i\,,\nonumber\\
f^3_{12}=1\,,&\quad&f^3_{03}=i\,.\nonumber
\end{eqnarray}
These triplets satisfy the standard algebra (\ref{algf}) and, in
addition, the unit roots of different triplets commute with each other,
\begin{equation}\label{comff}
\left[\,\left<f^i\right>\,,\,\left<f^j\right>^*\right]=0\,,\quad i,j=1,2,3.
\end{equation}
This property that holds in the Minkowski geometry seems to be particular since
this could not be proved so far in the general case.

The first triplet, ${\bf f}$,  gives rise to the spin-like operators
\[
\Sigma^i=\frac{1}{2}f^i_{\mu\nu}S^{\mu\nu}=\left(
\begin{array}{cc}
\sigma_i&0\\
0&0
\end{array}\right) \,,
\]
while the second one,  ${\bf f}^*$, yields
\[
\overline{\Sigma}^i=\frac{1}{2}\left(f^i_{\mu\nu}\right)^{*}S^{\mu\nu}
=\left(
\begin{array}{cc}
0&0\\
0&\sigma_i
\end{array}\right) \,.
\]
Because of the supplemental condition given by Eq. (\ref{comff})
the operators $\Sigma^i$ and $\overline{\Sigma}^i$ have a special form. 
They act {\em separately}
on orthogonal subspaces that are nothing other than the left ($\Psi_L$) and
respectively right-handed ($\Psi_R)$ chiral projections of the space of the 
Dirac spinors
$\Psi=\Psi_L \oplus \Psi_R$. Consequently, these subspaces will be the carrier 
spaces of
the irreducible representations  $spin({G}_{\bf f})$ and  
$spin({G}_{{\bf f}^*})$.
In addition, it is interesting to observe that 
$\Sigma^{(i)}+\overline{\Sigma}^i=2 S^i$.

The Dirac-type operators of the first triplet are
\begin{equation}\label{Dto}
D^i=i[D,\,\Sigma^i]=\left(
\begin{array}{cc}
0&-i\sigma_i D^{(+)}\\
iD^{(-)}\sigma_i&0
\end{array}\right) \,,
\end{equation}
while the second triplet gives us  $\overline{D}^i=i[D,\,\overline{\Sigma}^i]$.
The operators of each triplet and $D$ obey the anticommutation rules 
(\ref{4sup}) and
anticommute with $\gamma^0$. The operators $(D,D^i)$ form a basis for the 
D-superalgebra
$({\bf d}_{\bf f})_c$  while the conjugated algebra 
$({\bf d}_{{\bf f}^*})_c$ has
the basis $(D,\overline{D}^i)$.
It is remarkable that there are particular relations,
\[
\left\{ D^i\,,\,\overline{D}^j\right\}=-2\delta_{ij}D^2+4{\cal P}_i{\cal P}_j + 4i
\varepsilon_{ijk}{\cal P}_0{\cal P}_k
\]
involving operators from both different bases.

\subsection{Symmetries}

The next objective is to construct the groups of automorphisms of
the complex D-superalgebras. The first group is ${\rm Aut}({\bf
d}_{\bf f})_c= spin[(G_{\bf f})_c \,\circledS\, S(M_n)]$ where the
isometries act according to Eq. (\ref{UDUR}) with $\hat {\frak R}\in
O_{\bf f}\sim O_c(3)$. Our general results indicate that the
generators  of $spin (G_{\bf f})_c\sim SL(2, {\Bbb C})$ are  $\hat
s_i=\frac{1}{2}\Sigma^{(i)}$ and $\hat r_i
=\frac{i}{2}\Sigma^{(i)}$. These generators satisfy usual $sl(2,
{\Bbb C})$ commutation rules (\ref{sss}) and commute with the basis
generators of $spin[s(M)]$ as
\begin{eqnarray}
&[{\cal P}_{\mu},\,\hat s_i]=0\,,\quad [{\cal J}_i ,\, 
\hat s_j]=i\varepsilon_{ijk} \hat s_k\,,\quad
&[{\cal K}_i,\, \hat s_j]=i\varepsilon_{ijk} \hat r_k\,,\nonumber\\
&[{\cal P}_{\mu},\,\hat r_i]=0\,,\quad [{\cal J}_i,\, 
\hat r_j]=i\varepsilon_{ijk} \hat r_k\,,\quad
&[{\cal K}_i,\, \hat r_j]=i\varepsilon_{ijk} \hat s_k\,.\label{ubauba}
\end{eqnarray}
The transformations matrices $T(\rho)$ (with $\rho=\rho_if^{(i)}$ and 
$\rho_i = \rho'_i+
i\rho''_i$ where $ \rho'_i, \rho''_i \in {\Bbb R}$) that give the 
transformation laws
(\ref{TDT10}) and (\ref{TDT20})are
\[
T(\rho)=e^{-i\rho_i \Sigma^i}=e^{-2i(\rho'_i \hat s_i+\rho''_i \hat r_i)}\,.
\]
as it results from Eq. (\ref{TxiS}).
Hereby we observe that the matrices $T(\rho)$ act only on the
left-handed subspace $\Psi_L$ of the Dirac spinors.

The next step is to construct the complex-valued orthogonal matrices
of the group $O_{\bf f}\sim O_c(3)$ defined by Eq. (\ref{rotroot}). These have
the general form
\[
\hat{\frak R}_{ij}(\omega)=\textstyle\frac{1}{4}
f^{(i)\,\alpha\beta}
\Lambda^{\mu\,\cdot}_{\cdot\,\alpha}(\omega)
\Lambda^{\nu\,\cdot}_{\cdot\,\beta}(\omega)
f^{(j)}_{\mu\nu}\,,
\]
and constitute a representation of the group
$I(M)$  induced by the group $O(3)_c$. Of course, the translations have no 
effects in this
representation such that we are left only with the transformations  
$\Lambda(\omega)
\in O(3,1)$. These give rise to non-trivial matrices $\hat{\frak R}(\omega)$
as, for example,
\[
\hat{\frak R}(\varphi)=\left(
\begin{array}{ccc}
1&0&0\\
0&\cos\varphi&\sin\varphi\\
0&-\sin\varphi&\cos\varphi
\end{array}\right)\,,\quad
\hat{\frak R}(\alpha)=\left(
\begin{array}{ccc}
1&0&0\\
0&\cosh\alpha&i\sinh\alpha\\
0&-i\sinh\alpha&\cosh\alpha
\end{array}\right)\,,
\]
calculated for non-vanishing parameters $\omega_{23}=\varphi$ (a rotation
around $x^1$) and respectively $\omega_{01}=\alpha$ (a boost along $x^1$).
Thus we have all the ingredients we
need to write down the action of the transformations of the group
${\rm Aut}({\bf d}_{\bf f})_c$. We note that Eqs. (\ref{ubauba}) show that
$(G_{\bf f})_c$ is an invariant subgroup.
The second group of automorphisms, ${\rm Aut}({\bf d}_{{\bf f}^*})_c$, can be
constructed taking into account that the transformation matrices of
$spin(G_{{\bf f}^*})$ are $\overline{T}(\rho)$ and
the group $O_{{\bf f}^*}$ is formed by the conjugated matrices 
$\hat{\frak R}^*$.
We observe that  $\overline{T}$ are generated by
the matrices $\overline{\hat s}_i$ and  $\overline{\hat r}_i$  that act 
only on the right-handed subspace $\Psi_R$.

Hence it is clear that each chiral sector has its own set of unit roots
defining Dirac-type operators and
 groups of automorphisms of the D-super\-al\-ge\-bras for each chiral 
 sector separately.
The origin of this perfect balance between the chiral sectors is the form of
the operator $D^{(+)}D^{(-)}=D^{(-)}D^{(+)}$ that commutes with $\sigma_i$.

Finally we note that the principal problem that remains open
is the physical significance of the new Dirac-type operators of the Minkowski
spacetime we have studied here.

\subsection*{Acknowledgments}

This work is partially supported by MEC-CEEX D11-49/2005 Program, Romania.

\end{document}